\documentclass[amsmath,amssymb,noshowpacs,aip,pop,
preprint,
citeautoscript  
]{revtex4-1}
\usepackage{graphicx}
\usepackage{dcolumn}
\usepackage{bm}

\usepackage[utf8]{inputenc}
\usepackage[T1]{fontenc}
\usepackage{mathptmx}
\usepackage{etoolbox}

\makeatletter
\def\@email#1#2{%
 \endgroup
 \patchcmd{\titleblock@produce}
  {\frontmatter@RRAPformat}
  {\frontmatter@RRAPformat{\produce@RRAP{*#1\href{mailto:#2}{#2}}}\frontmatter@RRAPformat}
  {}{}
}%
\makeatother

\begin{document}

\title[GK EM Particle simulations in C1]{Gyrokinetic Electromagnetic Particle Simulations in Triangular Meshes with C1 Finite Elements}
\author{Zhixin Lu}
\email{{zhixin.lu@ipp.mpg.de}}

\author{Guo Meng}
\author{Roman Hatzky} 
\author{Eric Sonnendrücker}
\affiliation{Max Planck Institute for Plasma Physics, 85748 Garching, Germany}

\author{Alexey Mishchenko}
\affiliation{Max Planck Institute for Plasma Physics, 17491 Greifswald, Germany}

\author{Jin Chen}
\affiliation{Plasma Physics Laboratory, Princeton University, PO Box 451, Princeton, NJ 08543, 
United States of America}

\author{Philipp Lauber}
\affiliation{Max Planck Institute for Plasma Physics, 85748 Garching, Germany}

\author{Fulvio Zonca}
\affiliation{Center for Nonlinear Plasma Science; C.R. ENEA Frascati, C.P. 65, 00044 Frascati, Italy}

\author{Matthias Hoelzl}
\affiliation{Max Planck Institute for Plasma Physics, 85748 Garching, Germany}

\date{\today}

\begin{abstract}

The triangular mesh-based gyrokinetic scheme enables comprehensive axis-to-edge studies across the entire plasma volume. Our approach employs triangular finite elements with first-derivative continuity (C1), building on previous work to facilitate gyrokinetic simulations. Additionally, we have adopted the mixed variable/pullback scheme for gyrokinetic electromagnetic particle simulations. The filter-free treatment in the poloidal cross-section with triangular meshes introduces unique features and challenges compared to previous treatments using structured meshes. Our implementation has been validated through benchmarks using ITPA-TAE (Toroidicity-induced Alfv\'en Eigenmode) parameters, showing its capability in moderate to small electron skin depth regimes. Additional examinations using experimental parameters confirm its applicability to realistic plasma conditions.

\end{abstract}
  
\maketitle


\section{Introduction}
Since the early work of gyrokinetic particle-in-cell (PIC) simulations for the studies of tokamak plasmas \cite{lee1983gyrokinetic}, various methods have been developed  
for improving the simulation quality, especially for electromagnetic simulations, such as the $p_\parallel$ formulae and the iterative scheme for solving Amp\`ere's law \cite{chen2007electromagnetic,hatzky2019reduction}, the noisy matrix \cite{mishchenko2005gyrokinetic}, the conservative scheme \cite{bao2018conservative}, the mixed variable/pullback scheme \cite{mishchenko2019pullback,hatzky2019reduction,kleiber2024euterpe} and the implicit scheme \cite{lu2021development,sturdevant2021verification}.  The noise reduction scheme has been summarized comprehensively and analyzed numerically in the fully global linear gyrokinetic code GYGLES recently \cite{hatzky2019reduction}, which demonstrates the excellent performance of the iterative pullback method in noise reduction and the enhancement of simulation quality. 
The pullback scheme using mixed variables is implemented in {the global gyrokinetic Lagrangian Particle-In-Cell code}  ORB5 using the $\delta f$ scheme for the studies of EP-driven TAEs  \cite{mishchenko2019pullback}, demonstrating its capability in the MHD limit. Recently, the mixed variable/pullback scheme has also been applied to the full $f$ model for the studies of Alfv\'en waves and EP physics \cite{lu2023full}. 


While most previous gyrokinetic codes have been based on structured meshes and are dedicated to core plasmas \cite{wang2015distinct,chen2007electromagnetic,lin1998turbulent}, the unstructured meshes \cite{chang2017fast,lu2019development} or the multi-domain-structured meshes \cite{holzl2024non} have been used for whole plasma volume studies. Although the C1 triangular finite element has been applied to studies of tokamak plasmas using a fluid model \cite{jardin2004triangular}, the gyrokinetic particle codes are mainly based on C0 linear triangular finite elements \cite{chang2017fast,lu2019development},  where C0 indicates the zeroth-order continuity along the triangle edges, or the discontinuous Galerkin method \cite{jo2022development}. 
In this work, we develop the particle simulation scheme using C1 triangular finite elements, where C1 indicates the first-order continuity along the triangle edges. We focus on applying the noise reduction scheme to the simulations of drift waves, Alfv\'en waves, and energetic particle physics. 
By following the formulation from previous work \cite{hatzky2019reduction}, we implement the pullback scheme using mixed variables in the TRIMEG (TRIangular MEsh based Gyrokinetic) code \cite{lu2019development,lu2021development}. 
The TRIMEG code was initially developed using unstructured triangular meshes with C0 linear finite elements for whole plasma volume simulations of the electrostatic ion temperature gradient mode \cite{lu2019development}. 
It was later extended to study Alfv\'en waves and energetic particle physics using {the} mixed $\delta f$-full $f$ electromagnetic model in the testbed TRIMEG-GKX (Generalized Kinetics eXtended), which utilizes structured meshes \cite{lu2021development,lu2023full}. 
In the following, a high-order C1 finite element representation is developed in TRIMEG for upgrading the code to enable further studies including open field lines with high-order resolution of the field solver and particle-field coupling. 

This paper is organized as follows. In Sec. \ref{sec:model}, the equations for the discretization of the distribution function are derived with the mixed variables and the pullback scheme. In Sec. \ref{sec:numeric}, the normalized equations are given and the numerical methods are introduced. In Sec. \ref{sec:results}, various simulations of instabilities such as the energetic particle-driven toroidicity induced Alfv\'en eigenmode are performed, demonstrating the features of the schemes and key issues for the accurate description of the energetic particles in tokamak plasmas.

\section{Physics models and equations}
\label{sec:model}

\subsection{Discretization of distribution function}
The physics model in this work is closely connected to our previous work using a structured mesh \cite{lu2023full} and the references therein \cite{hatzky2019reduction,lanti2020orb5}. However, we focus on the traditional $\delta f$ model. Using the particle simulation scheme, $N$ markers are used with a given distribution,
\begin{align}
    g(z,t)\approx \sum_{p=1}^{N} \frac{\delta[z_p-z_p(t)]}{J_z}\;\;,
\end{align}
where $z$ is the phase space coordinate, $\delta$ is the Dirac delta function, $J_z$ is the corresponding Jacobian and $z=({\boldsymbol R},v_\|,\mu)$ is the phase space coordinate adopted in this work, $\mu=v_\perp^2/(2B)$ is the magnetic moment, $v_\|$ is the parallel velocity, ${\boldsymbol R}$ is the real space coordinate. The total distribution of particles is represented by the markers,
\begin{align}
    f(z,t)=C_{\rm{g2f}} P_{\rm{tot}}(z,t)g(z,t)
    \approx C_{\rm{g2f}} \sum_{p=1}^{N} p_{p,\rm{tot}}(t) \frac{\delta[z_p-z_p(t)]}{J_z}\;\;,
\end{align}
where the constant $C_{\rm{g2f}}\equiv N_f/N_g$, $N_{f/g}$ is the number of particles/markers, and $g$ and $f$ refer to the markers and physical particles respectively. 
For collisionless plasmas, 
\begin{align}
    \frac{{\rm d}g(z,t)}{{\rm d}t}=0 \;\;, \;\;
    \frac{{\rm d}f(z,t)}{{\rm d}t}=0 \;\;. 
\end{align}
For each marker, 
\begin{align}
    p_{p,\rm{tot}}(t)=\frac{1}{C_{\rm{g2f}}}\frac{f(z_p,t)}{g(z_p,t)}
    =\frac{N_g}{N_f}\frac{n_f}{n_g} \frac{f_v(z_p,t)}{g_v(z_p,t)}
    =\frac{n_f}{\langle n_f\rangle_V}\frac{\langle n_g \rangle_V}{n_g} \frac{f_v(z_p,t)}{g_v(z_p,t)} \;\;,
\end{align}
where  $n_f$ is the density profile and $f_v$ is the distribution in velocity space, namely, the particle distribution function $f=n_f({\boldsymbol R})f_v(v_\parallel,\mu)$, $\langle \ldots\rangle_V$ indicates the volume average. There are different choices of the marker distribution functions as discussed previously \cite{hatzky2019reduction,lanti2020orb5}. In this work, the markers are randomly initialized, uniformly distributed in the toroidal direction and in the $(R, Z)$ plane, while their velocity space distribution is identical to that of the physical particles, which leads to 
\begin{align}\label{eq:ptot_general}
    p_{p,\rm{tot}}(z,t) =\frac{\phi_{\rm wid}SR}{V_{\rm tot}}\;\;,
\end{align}
where $\phi_{\rm wid}$ is the width of the simulation domain in the toroidal direction, $S$ is the area of the poloidal cross-section, $V_{\rm tot}$ is the total volume. Equation~(\ref{eq:ptot_general}) is reduced to $p_{p,\rm{tot}}(z,t)={n_f}{R}/({\langle n_f\rangle_V}{R_0})$ for the tokamak equilibrium with concentric circular flux surfaces.

For the $\delta f$ model, the total distribution function is decomposed into the background and perturbed parts, $f(z,t)=f_0(z,t)+\delta f(z,t)$. The background part can be chosen as the time-independent one, i.e., $f_0(z,t)=f_0(z)$, and a typical choice is the local Maxwellian distribution. The background and perturbed distribution functions are represented by the markers as follows,
\begin{align}
    f_0(z,t)=P(z,t)g(z,t)\approx \sum_{p=1}^{N} p_p(t) \frac{\delta[z_p-z_p(t)]}{J_z}\;\;, \\
    \delta f(z,t)=W(z,t)g(z,t)\approx \sum_{p=1}^{N} w_p(t) \frac{\delta[z_p-z_p(t)]}{J_z}\;\;,
\end{align}
where $p_{p}(t)=f_0(z_p,t)/g(z_p,t)=P(z_p,t)$ and $w_{p}(t)=\delta f(z_p,t)/g(z_p,t)=W(z_p,t)$ are time-varying variables. The evolution equations are readily obtained \cite{lanti2020orb5},
\begin{align}
    &\dot w_p(t)  = - p_p(t) \frac{{\rm d}}{{\rm d}t} \left[\ln f_0(z_p(t)) \right] \;\;, \\
    &\dot p_p(t) =   p_p(t) \frac{{\rm d}}{{\rm d}t} \left[\ln f_0(z_p(t)) \right] \;\;, \\
    &\frac{{\rm d}}{{\rm d}t} = \frac{\partial}{\partial t} + \boldsymbol{\dot R}\cdot\nabla + \dot v_\|\frac{\partial}{\partial v_\|} \;\;,
\end{align}
where  the magnetic moment is assumed to be constant for the last equation, i.e., $\dot\mu= 0$. Generally, the gyro center's equation of motion can be decomposed into the equilibrium part corresponding to that in the equilibrium magnetic field, and the perturbed part due to the perturbed field 
\begin{eqnarray}
  {\boldsymbol{\dot R}} ={\boldsymbol{\dot R}}_0 +\delta {\boldsymbol{\dot R}} \;\;, \\
  \dot v_\| = \dot v_{\|,0}+\delta \dot{v} _\| \;\;.
\end{eqnarray}
The time derivative is defined as 
\begin{eqnarray}
    \frac{\rm d}{{\rm d}t}&=&\left.\frac{{\rm d}}{{\rm d}t}\right|_0 +\left.\frac{{\rm d}}{{\rm d}t}\right|_1 \;\;, \\
    \left.\frac{{\rm d}}{{\rm d}t}\right|_0&=& \frac{\partial}{\partial t} + {\boldsymbol{\dot R}}_0\cdot\nabla + \dot v_{\|,0} \frac{\partial}{\partial v_\|}  \;\;,\\
     \left.\frac{{\rm d}}{{\rm d}t}\right|_1&=&\delta {\boldsymbol{\dot R}} \cdot\nabla + \delta \dot v_{\|} \frac{\partial}{\partial v_\|} \;\;.
\end{eqnarray}
For the equilibrium distribution function,
\begin{eqnarray}
 \left.\frac{{\rm d}}{{\rm d}t}\right|_0f_0=0\;\;,
\end{eqnarray}
and thus
\begin{eqnarray}
 \frac{{\rm d}}{{\rm d}t} f_0 = \left.\frac{{\rm d}}{{\rm d}t}\right|_1f_0 \;\;,
\end{eqnarray}
where $f_0$ is chosen as a steady state solution ($\partial f_0/\partial t=0$).
In this work, the local Maxwell distribution is chosen ($f_0=f_{\rm{M}}$), 
\begin{eqnarray}
 f_{\rm{M}}=\frac{n_0}{(2T/m)^{3/2}\pi^{3/2}} \exp\left(-\frac{mv_\|^2}{2T}-\frac{m\mu B}{T}\right) \;\;,
\end{eqnarray}
with $\int f_{\rm{M}} J_v dv_\|d\mu=n_0,\;\; J_v=2\pi B$. And thus
\begin{eqnarray}
 \frac{{\rm d}}{{\rm d}t}\ln f_{\rm{M}} = \delta {\bf\dot R}\cdot\left[
 {\vec\kappa}_n + \left(\frac{mv_\|^2}{2T}+\frac{m\mu B}{T}-\frac{3}{2}\right)\vec\kappa_T
 -\frac{m\mu B}{T}\vec\kappa_B
 \right]
 - \delta \dot v_{\|} \frac{mv_\|}{T} \;\;, \nonumber\\
\end{eqnarray}
where $\vec\kappa_{n,T,B}\equiv \nabla\ln \{n,T,B\}$. 
Note that for the local Maxwell distribution, the following approximation has been made in the  $\delta f$ scheme in this work: ${\rm d}f_{\rm{M}}/{\rm d}t|_0=0$ is assumed to eliminate the neoclassical drive. Studies of neoclassical electron transport in TRIMEG(-C0) can be found in a separate work \cite{rekhviashvili2023gyrokinetic}. 

\subsection{Physics equations using mixed variables}
The mixed variable is defined as follows.
The parallel component $\delta A_\|$ of the scalar potential is decomposed into the symplectic part and the Hamiltonian part,
\begin{equation}
\label{eq:AsAh}
    \delta A_\| =\delta A_\|^{\rm{s}} + \delta A_\|^{\rm{h}} \;\;, 
\end{equation}
where the symplectic part is chosen to satisfy ideal Ohm's law involving the electrostatic scalar potential $\delta\Phi$ as follows,
\begin{equation}
    \partial_t\delta A_\|^{\rm{s}}+\partial_\|\delta\Phi = 0\;\;,
\label{eq:ohm_law0}
\end{equation}
where the parallel derivative is defined as $\partial_\|=\boldsymbol{b}_0\cdot\nabla$. 
The parallel velocity coordinate of the gyro center is defined as 
\begin{equation}
    u_\|=v_\|+\frac{q_s}{m_s}\langle\delta A_\|^{\rm{h}}\rangle\;\;,
\end{equation}
where $q_s$ and $m_s$ are the charge and mass of species $s$, respectively, the subscript $s$ represents the different particle species, and $\langle\ldots\rangle$ indicates the gyro average .

The gyro center's equations of motion are consistent with previous work \cite{mishchenko2019pullback,hatzky2019reduction,lanti2020orb5,mishchenko2023global,kleiber2024euterpe}, {
\begin{eqnarray}
 \dot{\boldsymbol R}_0 
  &=& u_\| {\boldsymbol b}^*_0 + \frac{m\mu}{qB^*_\|} {\boldsymbol b}\times\nabla B \;\;, 
  \\
  \dot u_{\|,0}
  &=& -\mu {\boldsymbol b}^*_0\cdot \nabla B \;\;,
  \\
  \delta\dot{\boldsymbol R}
  &=& \frac{{\boldsymbol b}}{B^*_\|}\times \nabla \langle \delta\Phi -u_\| \delta A_\|\rangle 
  -\frac{q_s}{m_s}\langle\delta A^{\rm h}_\|\rangle {\boldsymbol b}^*\;\;, 
  \\
  \delta \dot u_\|
  &=&  -\frac{q_s}{m_s} \left({\boldsymbol b}^*\cdot\nabla\langle\delta\Phi-u_\|\delta A^{\rm{h}}_\|\rangle +\partial_t\langle\delta A_\|^{\rm{s}}\rangle \right) \nonumber\\
  &&-\frac{\mu}{B^*_\|}{\boldsymbol b}\times\nabla B\cdot\nabla\langle\delta A_\|^{\rm{s}}\rangle \;\;,  
\end{eqnarray}
where ${\boldsymbol b}^*={\boldsymbol b}_0^*+\nabla\langle\delta A_\|^{\rm s}\rangle\times{\boldsymbol b}/B_\|^*$, ${\boldsymbol b}^*_0={\boldsymbol b}+(m_s/q_s)u_\|\nabla\times{\boldsymbol b}/B_\|^*$, ${\boldsymbol b}={\boldsymbol B}/B$, $B_\|^*=B+(m_s/q_s)u_\|{\boldsymbol b}\cdot(\nabla\times{\boldsymbol b})$. Note that in our previous work \cite{lu2023full}, $v_\|$ is adopted on the right-hand side, and  the term $-(q_s/m_s)\langle\delta A_\|^{\rm{h}}\rangle{\boldsymbol b}^*$ in $\dot{\boldsymbol R}$ is thus taken into account in $\dot{\boldsymbol R}_0$. 

The linearized quasi-neutrality equation in the long wavelength approximation is as follows, 
\begin{equation}
\label{eq:poisson0}
    -\nabla\cdot\left( \sum_s\frac{q_s n_{0s}}{B\omega_{{\rm c}s}} \nabla_\perp\delta\Phi \right) = \sum_s q_s \delta n_{s,v} \;\;,
\end{equation}
where the gyro density $\delta n_s$ is calculated using $\delta f_s({\boldsymbol R},v_\|,\mu)$ (indicated as $\delta f_{s,v}$), namely, $\delta n_s({\mathbf{x}})=\int {\rm d}^6 z\delta f_{s,v}\delta({\mathbf{R+\rho-x}})$, $\omega_{{\rm c}s}$ is the cyclotron frequency of species `$s$' and in this work, we ignore the perturbed electron polarization density on the left-hand side. 
When the $\delta f$ scheme is adopted, $\delta f_{s,v}$ is obtained from $\delta f_{s,u}$ as follows, with the linear approximation of the pullback scheme,
\begin{eqnarray}
    & \delta f_{s,v} = \delta f_{s,u} +  \frac{q_s\left\langle\delta A^{\rm{h}}_{\|} \right\rangle}{m_s}\frac{\partial f_{0s}}{\partial v_\|}
    \xrightarrow[f_{0s}=f_{\rm{M}}]{\text{Maxwellian}}
     \delta f_{s,u} -  \frac{ v_\|}{T_s}  q_s\left\langle\delta A^{\rm{h}}_{\|} \right\rangle f_{0s} \;\;,
\end{eqnarray}
which is obtained from the more general form of the nonlinear pullback scheme \cite{hatzky2019reduction},
\begin{eqnarray}
    & f_{s,v} (v_\|) = f_{s,u} \left(v_\|+\frac{q_s}{m_s}\langle\delta A_\|^{\rm h}\rangle\right)\;\;.
\end{eqnarray}

Amp\`ere's law in $v_\|$ space is given by 
\begin{equation}
    -\nabla^2_\perp\delta A_\| = \mu_0 \delta j_{\|,v} \;\;,
\end{equation}
where $\delta j_{\|,v}({\mathbf{x}})=\sum_s q_s \int {\rm d}^6 z\delta f_{v,s}\delta({\mathbf{R+\rho-x}})v_\|$.

For the $\delta f$ model, using the mixed variables and assuming a Maxwell distribution, we have
\begin{eqnarray}
    \delta j_{\|,v}
    &\equiv&\sum_s q_s\int \mathrm{d}z^6\delta f_{s,v}(v_\|)\delta({\mathbf{R+\rho-x}}) v_\| \nonumber \\
    &=&\sum_s q_s\int \mathrm{d}z^6 \left(\delta f_{s,u}(u_\|) -\frac{v_\| q_s\langle \delta A_\|^{\rm{h}}\rangle }{T_s}f_{0s}\right) \delta({\mathbf{R+\rho-x}})  v_\| \;\;.
\end{eqnarray}
Then we can write Amp\`ere's law as {
\begin{eqnarray}
\label{eq:ampere_mv_deltaf_exact}
    -\nabla^2_\perp\delta A_{\|}^{\rm{h}}
    +\sum_s\mu_0\frac{q_s^2}{T_s}\int \mathrm{d}z^6 v_\|^2 f_{0s} \langle \delta A_{\|}^{\rm{h}} \rangle \delta({\mathbf{R+\rho-x}}) \nonumber\\
    =\nabla^2_\perp\delta A_{\|}^{\rm{s}} 
    + \mu_0\sum_s q_s  \int \mathrm{d}z^6  v_\| \delta f_{s,u}(u_\|)\delta({\mathbf{R+\rho-x}})   \;\;.
\end{eqnarray}}
The integral on the left-hand side can be obtained analytically, yielding
\begin{eqnarray}
\label{eq:ampere_mv_deltaf}
    -\nabla^2_\perp\delta A_{\|}^{\rm{h}}
    +\sum_s\frac{1}{d_{s}^2}\overline{\langle\delta A_{\|}^{\rm{h}}\rangle  }
    &=&\nabla^2_\perp\delta A_{\|}^{\rm{s}} 
     \nonumber \\
     &+&\mu_0\sum_s q_s  \int \mathrm{d}z^6  v_\| \delta f_{s,u}(u_\|)  \delta({\mathbf{R+\rho-x}}) \;\;, 
\end{eqnarray}
with
\begin{eqnarray}
\label{eq:dAh_from_f0}
    \overline{\langle\delta A_{\|}^{\rm{h}}\rangle  }
    &\equiv&\frac{2}{n_{s0}v_{ts}^2}\int \mathrm{d}z^6 v_\|^2 f_{0s} \langle \delta A_{\|}^{\rm{h}} \rangle \delta({\mathbf{R+\rho-x}}) \,\,,
\end{eqnarray}
where $v_{\rm{t}s}=\sqrt{2T_s/m_s}$, $d_{s}$ is the skin depth of species `$s$' defined as $d_{s}^2=c^2/\omega_{p,s}^2=m_s/(\mu_0q_s^2n_{0s})$ and the integral in Eq.~(\ref{eq:dAh_from_f0}) is kept without analytical reduction in order to capture the numerical or physics deviation of $f_0$ away from the Maxwellian distribution. 

For the full $f$ model, the perturbed current is represented by the full $f$,
\begin{eqnarray}
    \delta j_{\|,v}&=&\sum_s q_s\int \mathrm{d}v^3 v_\| f_{s,v}  \nonumber \\
    &=&  \sum_sq_s\int \mathrm{d}z^6 \left[ u_\|-\frac{q_s}{m_s}\langle\delta A_\|^{\rm{h}}\rangle\right] f_{s,v} \delta({\mathbf{R+\rho-x}})\;\;. 
\end{eqnarray} 
Amp\`ere's law yields {
\begin{eqnarray}
    -\nabla^2_\perp\delta A_{\|}^{\rm{h}}
    +\mu_0\sum_s\frac{q_s^2}{m_s}\int \mathrm{d}z^6 f_{s,v}\langle \delta A_{\|}^{\rm{h}} \rangle\delta({\mathbf{R+\rho-x}})  \nonumber \\
    =\nabla^2_\perp\delta A_{\|}^{\rm{s}} + \mu_0\sum_s q_s  \int \mathrm{d}z^6  u_\| f_{s,v} \delta({\mathbf{R+\rho-x}}) \;\;.
\end{eqnarray}}
The corresponding analytical limit gives the similar form of Eq.~(\ref{eq:ampere_mv_deltaf}) except the replacement of $\delta f_{s,u}(u_\|)$ with $f_{s,v}(v_\|)$ and the definition of $\overline{\langle\delta A_{\|}^{\rm{h}}\rangle  }$,
\begin{eqnarray}
    -\nabla^2_\perp\delta A_{\|}^{\rm{h}}
    +\sum_s\frac{1}{d_{s}^2}\overline{\langle\delta A_{\|}^{\rm{h}}\rangle  }& =&
    \nabla^2_\perp\delta A_{\|}^{\rm{s}} 
     \nonumber \\
    &+&\mu_0\sum_s q_s  \int \mathrm{d}z^6  v_\| f_{s,v}(v_\|)  \delta({\mathbf{R+\rho-x}})  \;\;, 
\end{eqnarray}
with 
\begin{eqnarray}
    \overline{\langle\delta A_{\|}^{\rm{h}}\rangle  }
    &\equiv&
    \frac{1}{n_{s0}}\int \mathrm{d}z^6  f_{s,v} (v_\parallel) \langle \delta A_{\|}^{\rm{h}} \rangle \delta({\mathbf{R+\rho-x}}) \;\;.
\end{eqnarray}

For both the $\delta f$ model and the full $f$ model, using the iterative scheme, the asymptotic solution is expressed as follows,
\begin{equation}
    \delta A^{\rm{h}}_{\|}=\sum_{I=0}^\infty\delta A^{\rm{h}}_{\|,I}\;\;,
\end{equation}
where $|\delta A^{\rm{h}}_{\|,I+1}/\delta A^{\rm{h}}_{\|,I}|\ll1$ assured by the fact that the analytical skin depth term (the second term on the left-hand side of Eq.~(\ref{eq:ampere_mv_deltaf})) is close to the exact one (the second term on the right-hand side of Eq.~(\ref{eq:ampere_mv_deltaf})). 
Amp\`ere's law is solved order by order,
\begin{eqnarray}
\label{eq:ampere_h0}
    \left(\nabla^2_\perp-\sum_s\frac{1}{d_{s}^2}\right)\delta A_{\|,0}^{\rm{h}} 
    = -\nabla^2_\perp\delta A_{\|}^{\rm{s}} - \mu_0 \delta j_{\|} \;\;, \\
\label{eq:ampere_iterative}
    \left(\nabla^2_\perp-\sum_s\frac{1}{d_{s}^2}\right)\delta A_{\|,I}^{\rm{h}} 
    =-\sum_s\frac{1}{d_{s}^2}\delta A^{\rm{h}}_{\|,I-1} 
    + \sum_s\frac{1}{d_{s}^2} \overline{\langle\delta A_{\|,I-1}^{\rm{h}}\rangle}\;\;, \\
    \overline{\langle\delta A_{\|,I-1}^{\rm{h}}\rangle}
    =\frac{2}{n_0 v_{\rm{t}s}^2}\int \mathrm{d}z^6 v_\|^2 f_{0s}  \langle\delta A^{\rm{h}}_{\|,I-1} \rangle\delta({\mathbf{R+\rho-x}})  \;\;,\text{ for $\delta f$ model} \\
\label{eq:A2ndavg_fullf}
    \overline{\langle\delta A_{\|,I-1}^{\rm{h}}\rangle}
    =\frac{1}{n_0}\int \mathrm{d}z^6 f_{s,v}  \langle\delta A^{\rm{h}}_{\|,I-1} \rangle \delta({\mathbf{R+\rho-x}}) \;\;,\text{ for full $f$ model}
\end{eqnarray}
where  $I=1,2,3,\ldots$. Note that since $2/(n_0 v_{\rm{t}s}^2)\int \mathrm{d}v^3 v_\|^2 f_0=1$ and {$(1/n_0)\int \mathrm{d}v^3  f_0=1$} for the Maxwell distribution in the analytical limit by ignoring the finite Larmor radius effect (which is well satisfied for electrons, the main contributor of the skin depth term), namely, the analytical electron skin depth term is close to the exact one, and thus good convergence of the iterative solver is expected for the efficient and accurate calculation of the skin depth term.

\subsection{Pullback scheme for mitigating the cancellation problem}
A more detailed description of the pullback scheme can be found in the previous work \cite{mishchenko2019pullback}. As a brief review, the equations for the $\delta f$ are listed as follows.
\begin{align}
\label{eq:pullback_A}
    & \delta A^{\rm{s}}_{\|,\rm{new}} = \delta A^{\rm{s}}_{\|,\rm{old}} + \delta A^{\rm{h}}_{\|,\rm{old}}  \;\;, \\
\label{eq:pullback_v}
    & u_{\|,\rm{new}} = u_{\|,\rm{old}} - \frac{q_s}{m_s} \left\langle\delta A^{\rm{h}}_{\|,\rm{old}} \right\rangle  \;\;, \\
\label{eq:pullback_df}
    & \delta f_{\rm{new}} = \delta f_{\rm{old}} +  \frac{q_s\left\langle\delta A^{\rm{h}}_{\|,\rm{old}} \right\rangle}{m_s}\frac{\partial f_{0s}}{\partial v_\|}
    \xrightarrow[f_{0s}=f_{M}]{\text{Maxwellian}}
     \delta f_{\rm{old}} -  \frac{ 2v_\|}{v_{\rm{t}s}^2}  \frac{q_s\left\langle\delta A^{\rm{h}}_{\|,\rm{old}} \right\rangle}{m_s} f_{0s} \;\;,
\end{align}
where variables with subscripts `new' and `old' refer to those after and before the pullback transformation, Eq.~(\ref{eq:pullback_df}) is the linearized equation for $\delta f$ pullback, which is from the general equation of the transformation for the distribution function 
\begin{align}
    f_{\rm{old}} (u_{\| \rm{old}}) = f_{\rm{new}} \left(u_{\| \rm{new}} =u_{\| \rm{old}}- \frac{q_s}{m_s} \left\langle\delta A^{\rm{h}}_{\|,\rm{old}} \right\rangle \right) \;\;. 
\end{align}
For the full $f$ scheme, only Eqs.~(\ref{eq:pullback_A}) and (\ref{eq:pullback_v}) are needed.

\section{Numerical methods}
\label{sec:numeric}

\subsection{Normalized equations}
The normalization of the variables in the TRIMEG-C1 code is introduced in this section. The reference length for normalization is $R_{\rm{N}}=1$ m since not only the micro-instabilities but also the large-scale modes such as Geodesic acoustic mode are of interest in the future. In addition, using $R_{\rm N}=1$ m, the length variables in the equilibrium (EQDSK) files can be used in TRIMEG-C1 directly without further normalization. The particle mass is normalized to $m_{\rm{N}}$ and $m_{\rm{N}}=m_{\rm{p}}$ (the proton mass). 
The velocity unit is 
\begin{align*}
    v_{\rm{N}}\equiv \sqrt{2T_{\rm{N}}/m_{\rm{N}}} \;\;,
\end{align*}
where $T_{\rm{N}}$ and $m_{\rm{N}}$ are the reference temperature and mass for normalization. The time unit is $t_{\rm N}=R_{\rm N}/v_{\rm N}$. The charge unit is the elementary charge $e$. 
The temperature is normalized to $T_{\rm{N}}=m_{\rm{N}}v_{\rm{N}}^2/2$.
In addition, $\mu$ is normalized to $v_{\rm{N}}^2/B_{\rm N}$, 
\begin{align*}
    \mu\equiv\frac{ v_\perp^2}{2B}=\bar{\mu} \frac{v_{\rm{N}}^2}{B_{\rm N}}=\bar\mu\frac{2T_{\rm N}}{B_{\rm N}m_{\rm N}},
\end{align*}
where $B_{\rm N}=1\;{\rm T}$ in this work so that the magnetic field data from the EQDSK file can be used without further normalization. 

In addition to the units for the normalzation, two reference variables are defined, namely, $n_{\rm ref}$ and $B_{\rm ref}$, so that the two basic parameters $\beta_{\rm ref}$ and $\rho_{\rm N}$ can be calculated from $n_{\rm ref}$ and $B_{\rm ref}$ as follows,
\begin{eqnarray}
    \beta_{\rm ref}=\frac{\mu_0 n_{\rm ref} m_{\rm N} v_{\rm N}^2}{B_{\rm ref}^2} \;\;, \\
    \rho_{\rm N}=\frac{m_{\rm N}v_{\rm N}}{eB_{\rm ref}} \;\;,
\end{eqnarray}
where $\beta_{\rm ref}$ appears in the normalized Amp\`ere's law Eqs.~(\ref{eq:ampere_normalize0})--(\ref{eq:skin_fullf_normalize}) and $\rho_{\rm N}$ appears in the normalized gyro center's equation of motion Eqs.~(\ref{eq:dR0dt_normalized})--(\ref{eq:du1dt_normalized}).

The Maxwell distribution is 
\begin{eqnarray}
  f_{\rm{M}}
  = \frac{\bar n_0}{\bar v_{\mathrm t}^3\pi^{3/2}} e^{-\frac{mv_\|^2+2\mu B}{2T}}
  = \frac{\bar n_0}{\bar v_{\mathrm t}^3\pi^{3/2}} \exp \left(-\frac{\bar{m}\bar{v}^2_\|}{\bar{T}} -2\frac{\bar{m}\bar{\mu}_\|}{\bar{T}} \bar B \right) 
  \;\;,
\end{eqnarray}
and correspondingly,
\begin{eqnarray}
 \frac{{\rm d}}{{\rm d}\bar{t}}\ln f_{\rm{M}}& = &\delta \dot{\bf\bar R}\cdot\left[
 {\vec\kappa}_n + 
 \left(\frac{\bar m \bar v_\|^2}{\bar T}+\frac{2\bar m \bar\mu \bar B}{\bar T}-\frac{3}{2}\right)\vec\kappa_T
 -2\frac{\bar m\bar\mu \bar B}{\bar T}\vec\kappa_B
 \right] \nonumber\\
 &-& \frac{2\bar m\bar v_\|}{\bar T}  \frac{{\rm d}\delta \bar v_\|}{{\rm d}\bar{t}}\;\;.
\end{eqnarray}
The markers are loaded with the same distribution of physical particles in velocity space but uniformly in the poloidal plane and in the toroidal direction. In the $v_\|$ direction, a random number generator is used to produce numbers with normal distribution $f(x)=1/(\sigma\sqrt{2\pi})\exp\{-[(x-x_0)/\sigma]^2/2\}$, where $x_0$ and $\sigma$ are chosen as $0$ and $\sqrt{\bar{T}/(2\bar{m})}$ respectively.
In the $\mu$ direction, random numbers~$x$ uniformly distributed on $[0,1]$ are generated and shifted according to $\mu=-\ln(x)\bar{T}B_{\rm N}/(2\bar{m}B)$. 

The normalized gyro center's equations of motion are {
\begin{eqnarray}
\label{eq:dR0dt_normalized}
  \frac{{\rm d}{\Bar{\boldsymbol{R}}_0}}{{\rm d}\Bar{t}} 
  &=& \Bar{u}_\| {\boldsymbol b}^*_0 + \frac{\Bar{m}_s}{\Bar{q}_s}\Bar{\rho}_{\rm{N}}\frac{B_{\rm{ref}}}{B^2B^*_\|} \bar\mu{\boldsymbol B}\times\nabla B \;\;, 
  \\
\label{eq:du0dt_normalized}
  \frac{{\rm d}\bar u_{\|,0}}{{\rm d}\bar t} 
  &=& -\frac{\bar\mu}{B_{\rm N}} {\boldsymbol b}^*\cdot \bar\nabla B \;\;,
  \\
\label{eq:d1Rdt_normalized}
  \frac{{\rm d}\delta\bar{\boldsymbol R}}{{\rm d}\bar t} 
  &=& \Bar{\rho}_{\rm{N}}\frac{B_{\rm{ref}}}{B^*_\|}{\boldsymbol b}\times\bar \nabla \langle \delta\bar\Phi -\bar u_\| \delta \bar A_\|\rangle 
  -\frac{\bar q_s}{\bar m_s}\langle\delta \bar A^{\rm h}_\|\rangle {\boldsymbol b}^*\;\;, 
  \\
\label{eq:du1dt_normalized}
  \frac{{\rm d}\delta\bar u_\|}{{\rm d}\bar t} 
  &=&  -\frac{\Bar{q}_s}{\Bar{m}_s} \left({\boldsymbol b}^*\cdot\bar\nabla\langle\delta\bar\Phi-\bar u_\|\delta \bar A^{\rm{h}}_\|\rangle +\partial_{\bar t}\langle\delta \bar A_\|^{\rm{s}}\rangle \right)  \nonumber\\ &-&\bar\rho_{\rm{N}}\frac{B_{\rm ref}}{B^*_\|B_{\rm N}}\bar\mu{\boldsymbol b}\times\bar\nabla B\cdot\bar\nabla\langle\delta\bar A_\|^{\rm{s}}\rangle \;\;,
\end{eqnarray}
where  $\bar\nabla=R_{\rm N}\nabla$,} $\delta \Phi$ and $\delta A_\|$ are normalized to $m_{\rm{N}} v_{\rm{N}}^2/e$ and $m_{\rm{N}} v_{\rm{N}}/e$ respectively. 

The normalized quasi-neutrality equation is, {
\begin{eqnarray}
\label{eq:poisson_normalized}
    \bar\nabla_\perp \cdot \left(G_{\rm P}\bar\nabla_\perp\delta \bar{\Phi}\right)=-C_{\rm P}\delta\bar N\;\;, \;\;
    G_{\rm P}=\sum_s \frac{n_{0s}}{n_{\rm ref}} \bar{m}_s \left(\frac{B_{\rm N}}{B}\right)^2 \;\;,
\end{eqnarray} }
where $C_{\rm P}=1/\bar\rho_{\rm{N}}^2$, $\delta\bar{N}_s=\delta n/n_{\rm ref}$, $\delta\bar{N}=\sum_s\bar q_s\delta\bar{N}_s$.

For Amp\`ere's law, the original normalized equation $\nabla^2_\perp\delta \bar A =C_{\mathrm A} \delta\bar J_{\|,v}$ is solved using mixed variables and the iterative scheme corresponding to Eqs.~(\ref{eq:ampere_h0})--(\ref{eq:A2ndavg_fullf}),  {
\begin{eqnarray}
\label{eq:ampere_normalize0}
    &&\left(\bar\nabla^2_\perp-\sum_s\frac{\bar{q}_s^2}{\bar m_s}C_{\mathrm A}\right)\delta \bar A_{\|,0}^{\rm{h}} 
    = -\bar\nabla^2_\perp \delta\bar A_{\|,0}^{\rm{s}} - C_{\mathrm A} \delta \bar J_{\|} \;\;, 
    \\
\label{eq:ampere_normalize1}
    &&\left(\bar\nabla^2_\perp-\sum_s\frac{\bar{q}_s^2}{\bar m_s}C_{\mathrm A}\right)\delta\bar A_{\|,I}^{\rm{h}} 
    = -\sum_s\frac{\bar{q}_s^2}{\bar m_s}C_{\mathrm A}\delta\bar A_{\|,I-1}^{\rm{h}} 
    + \bar{G} \delta\bar A_{\|,I-1}^{\rm{h}} 
    \\
\label{eq:skin_df_normalize}
    &&\bar{G}\delta\bar A_{\|,I-1}^{\rm{h}} =C_{\mathrm A}\frac{N_{0s} \bar{q}_s^2}{\bar{T}_s}\sum_{p=1}^N 2\bar v_{\parallel,p}^2 
    \int {\rm d}z^6 w_p\delta(\Tilde{\mathbf{R}}_p)\langle\delta\bar A_{\|,I-1}^{\rm{h}}\rangle   {\text{ for $\delta f$}}\;\;,
    \\
\label{eq:skin_fullf_normalize}
    &&\bar{G}\delta\bar A_{\|,I-1}^{\rm{h}} =C_{\mathrm A}\frac{N_{0s} \bar{q}_s^2}{\bar m_s}\sum_{p=1}^N 
    \int {\rm d}z^6 p_{p,{\rm tot}}\delta(\Tilde{\mathbf{R}}_p) 
    \langle\delta\bar A_{\|,I-1}^{\rm{h}}\rangle   {\text{ for full $f$}} \;\;,
\end{eqnarray} }
where {$\Tilde{\boldsymbol{R}}_p={\boldsymbol{R}_p+\mathbf{\rho}_p-\mathbf{x}}$, $\mathbf{\rho}_p=m_s\mathbf{v}_{\perp,p}/(q_sB)$, $\mathbf{x}$ indicates the particle location,} $C_{\mathrm A} =\beta_{\rm ref}/\rho_{\rm N}^2$, $n_{\rm ref}$ is the reference density and is chosen as the electron density on magnetic axis, and $\delta\Bar{J}_\|=\delta j_\|/(ev_{\rm N} n_{\rm ref})$.

The normalized equations for the pullback treatment are as follows,{
\begin{align}
    & \delta\bar A^{\rm{s}}_{\|,\rm{new}} = \delta\bar A^{\rm{s}}_{\|,\rm{old}} + \delta\bar A^{\rm{h}}_{\|,\rm{old}}  \;\;, \\
    & u_{\|,\rm{new}} = \bar u_{\|,\rm{old}} - \frac{\bar{q}}{\bar{m}_s} \left\langle\delta\bar A^{\rm{h}}_{\|,\rm{old}} \right\rangle  \;\;, \\
    & \delta f_{\rm{new}} = \delta f_{\rm{old}} +  \frac{\Bar{q}_s\left\langle\delta\bar A^{\rm{h}}_{\|,\rm{old}} \right\rangle}{\bar{m}_s}\frac{\partial f_{0s}}{\partial \bar{v}_\|} \\
    \label{eq:pullback_df_norm}
    &
    \xrightarrow[f_{0s}=f_{\rm{M}}]{\text{Maxwellian}}
     \delta f_{\rm{old}} -  \frac{ 2 \bar{q}_s}{\bar{T}_s} \bar{v}_\| \left\langle\delta\bar A^{\rm{h}}_{\|,\rm{old}} \right\rangle f_{0s} \;\;,
\end{align}}
where the factor $2$ is from the normalization of $T$ to $T_{\rm{N}}=m_{\rm{N}}v_{\rm{N}}^2/2$, and Eq.~(\ref{eq:pullback_df_norm}) is the linearized pullback scheme for the $\delta f$ model implemented in our work. The studies using the nonlinear pullback scheme are beyond the scope of this work and will be addressed in the future. 

\subsection{Coordinates and input files}
The right-handed cylindrical coordinates $(R,\phi,Z)$ are adopted in TRIMEG-C1. The magnetic equilibrium is constructed using the standard EFIT output (the EQDSK ``g'' file), which provides the macroscopic parameters of the equilibrium such as the on-axis magnetic field $B_{\rm axis}$, the poloidal magnetic flux functions at the axis and the last closed surface; the one-dimensional profiles such as the safety factor $q$ and the toroidal current function $T$; and the two-dimensional poloidal magnetic flux function as a function of $(R,Z)$. The density and the temperature of any given species, such as the background electrons are given as functions of $R$ and $Z$ given in the input files to the TRIMEG-C1 code. External Matlab scripts are available to convert the one-dimensional  profiles (as functions of the poloidal magnetic flux) of the density and the temperature to two-dimensional profiles. The gyro centers' equations of motion are implemented in $(R,\phi,Z)$ coordinates, as described in  \ref{subsec:gcmotion_RZ_complete} and  \ref{subsec:gcmotion_RZ_simplified}.

\subsection{C1 finite element in triangular meshes}
The three-dimensional solver in this work is developed with the finite element method adopted in all directions in the $(R, \phi, Z)$ coordinate system.  The unstructured triangular mesh is adopted in the $(R,Z)$ directions while a uniform structured mesh is adopted in the $\phi$ direction.
The periodic boundary condition is adopted in the toroidal direction. In the $(R,Z)$ poloidal cross-section, a Dirichlet boundary condition with zero value of the function is implemented at the outer boundary of the simulation domain. 
The grid size is determined by the number of vertices in the $(R,Z)$ plane and the number of grid points $N_\phi$ in the $\phi$ direction.

Local coordinates $(\xi,\eta)$ are defined in each grid cell triangle in the poloidal plane. The three vertices of each triangle are located at $(\xi,\eta)=(0,0), (1,0), (0,1)$ as shown in Fig. \ref{fig:triangle}. For the C0 linear finite elements, three basis functions are adopted as follows,
\begin{eqnarray}
    \Lambda_1^{\rm lin} = 1-\xi-\eta\;\;, \\
    \Lambda_2^{\rm lin} = \xi\;\;,\\
    \Lambda_3^{\rm lin} = \eta\;\;.
\end{eqnarray}
The C1 finite element basis functions used in this work are similar to those in the GTS code  (the physics applications can be found in previous articles \cite{wang2015distinct,wang2015identification,lu2015intrinsic,lu2015effects}). Note that C1 FEM has been implemented in GTS but for productive runs, C1 has not been used previously since other solvers are sufficient for the electrostatic gyrokinetic simulations. The GTS C1 field solver is in $(\bar{r},\bar\theta)$ coordinates where $\bar{r}$ and $\bar\theta$ are the radial-like and poloidal-like coordinates, respectively, which is dedicated to core plasma studies. In this work, we developed the C1 scheme in the $(R,Z)$ coordinates in the newly written modules in TRIMEG-C1 so that it is possible to simulate the plasma with the open field line region in the future.   
The 18 basis functions of the C1 finite element method are given by \cite{jardin2004triangular}
\begin{eqnarray}
&&\Lambda_1=\lambda^2[10\lambda-15\lambda^2+6\lambda^3+30\xi\eta(\xi+\eta)] \;\;,\nonumber\\
&&\Lambda_2=\xi\lambda^2(3-2\lambda-3\xi^2+6\xi\eta) \;\;,\nonumber\\
&&\Lambda_3=\eta\lambda^2(3-2\lambda-3\eta^2+6\xi\eta) \;\;,\nonumber\\
&&\Lambda_4=0.5\xi^2\lambda^2(1-\xi+2\eta) \;\;,\nonumber\\
&&\Lambda_5=\xi\eta\lambda^3 \;\;, \nonumber\\
&&\Lambda_6=0.5\eta^2\lambda^2(1+2\xi-\eta) \;\; \nonumber,
\end{eqnarray}
\begin{eqnarray}
&&\Lambda_7=\xi^2(10\xi-15\xi^2+6\xi^3+15\eta^2\lambda) \;\;,\nonumber\\
&&\Lambda_8=0.5\xi^2(-8\xi+14\xi^2-6\xi^3-15\eta^2\lambda) \;\;,\nonumber\\
&&\Lambda_9=0.5\xi^2\eta(6-4\xi-3\eta-3\eta^2+3\xi\eta) \;\;, \\
&&\Lambda_{10}=0.25\xi^2[2\xi(1-\xi)^2+5\eta^2\lambda] \;\;,\nonumber\\
&&\Lambda_{11}=0.5\xi^2\eta(-2+2\xi+\eta+\eta^2-\xi\eta) \;\;,\nonumber\\
&&\Lambda_{12}=0.25\xi^2\eta^2\lambda+0.5\xi^3\eta^2 \;\;,\nonumber
\end{eqnarray}
\begin{eqnarray}
&&\Lambda_{13}=\eta^2(10\eta-15\eta^2+6\eta^3+15\xi^2\lambda) \;\;,\nonumber\\
&&\Lambda_{14}=0.5\xi\eta^2(6-3\xi-4\eta-3\xi^2+3\xi\eta) \;\;,\nonumber\\
&&\Lambda_{15}=0.5\eta^2(-8\eta+14\eta^2-6\eta^3-15\xi^2\lambda) \;\;,\nonumber\\
&&\Lambda_{16}=0.25\xi^2\eta^2\lambda+0.5\xi^2\eta^3 \;\;,\nonumber\\
&&\Lambda_{17}=0.5\xi\eta^2(-2+\xi+2\eta+\xi^2-\xi\eta) \;\;,\nonumber\\
&&\Lambda_{18}=0.25\eta^2[2\eta(1-\eta)^2+5\xi^2\lambda] \;\;, \nonumber
\end{eqnarray}
where $\lambda=1-\eta-\xi$.
In $(R,Z)$ coordinates, any variable $y(R,Z)$ can be written as
\begin{eqnarray}
    y(R,Z) &=&  \sum_{k=1}^{18}\sum_{j=1}^{N_{\rm tri}} \Lambda_{j,k}(\xi,\eta) Y_{j,k}  
               =\sum_{i=1}^{18N_{\rm tri}} \Lambda_i(\xi,\eta) Y_i  \;\;,
\end{eqnarray}
where $N_{\rm tri}$ is the number of the triangles and for the sake of simplicity, we converted the summation over $j$ and $k$ to that over $i$ according to $i=18(j-1)+k$. 
The coefficients $Y_i$ are solved with the constraint that the variable $y(R,Z)$, $\partial y(R,Z)/\partial R$, $\partial y(R,Z)/\partial Z$,  $\partial^2 y(R,Z)/\partial R^2$, $\partial^2 y(R,Z)/\partial R\partial Z$ and $\partial^2 y(R,Z)/\partial Z^2$ are continuous at any triangle vertex (C2). It can be shown that along the triangle edges, $y(R,Z)$ is C1 continuous \cite{jardin2004triangular,strang1971analysis}. The C2 continuity at any triangle vertex can be assured by switching $N_i$ to the set of basis functions $\hat{\Lambda}_i$ which are the linear combination of $\Lambda_i$,
\begin{eqnarray}
	y(R,Z) &=& \sum_{k=1}^{6}\sum_{j=1}^{N_{\rm vert}} \hat\Lambda_{j,k}(\xi,\eta)\hat Y_{j,k}  =\sum_{i=1}^{6N_{\rm vert}} \hat\Lambda_i(\xi,\eta) \hat{Y}_i  \;\;,
\end{eqnarray}
where $N_{\rm vert}$ is the number of vertices, `6' is the degree of freedom on each vertex except for the boundary vertices, and we converted the summation over $j$ and $k$ to that over $i$ according to $i=6(j-1)+k$. The basis functions $\hat\Lambda_i(\xi,\eta)$ is obtained as follows. Note that
\begin{eqnarray}
&&	\hat{Y}_i = \nonumber\\
&&	 \left\{
		\underbrace{y({\boldsymbol R}_1), 
			\frac{\partial y({\boldsymbol R}_1)}{\partial R}, 
			\frac{\partial y({\boldsymbol R}_1)}{\partial Z},
			\frac{\partial^2 y({\boldsymbol R}_1)}{\partial R^2},  
			\frac{\partial^2 y({\boldsymbol R}_1)}{\partial R\partial Z},
			\frac{\partial^2 y({\boldsymbol R}_1)}{\partial Z^2},
		}_{\rm vertex 1} \right. \nonumber \\
&&	\underbrace{y({\boldsymbol R}_2), 
		\frac{\partial y({\boldsymbol R}_2)}{\partial R}, 
		\frac{\partial y({\boldsymbol R}_2)}{\partial Z},
		\frac{\partial^2 y({\boldsymbol R}_2)}{\partial R^2},  
		\frac{\partial^2 y({\boldsymbol R}_2)}{\partial R\partial Z},
		\frac{\partial^2 y({\boldsymbol R}_2)}{\partial Z^2},
	}_{\rm vertex 2} \nonumber \\
&&
\left.
	\underbrace{y({\boldsymbol R}_3), 
	\frac{\partial y({\boldsymbol R}_3)}{\partial R}, 
	\frac{\partial y({\boldsymbol R}_3)}{\partial Z},
	\frac{\partial^2 y({\boldsymbol R}_3)}{\partial R^2},  
	\frac{\partial^2 y({\boldsymbol R}_3)}{\partial R\partial Z},
	\frac{\partial^2 y({\boldsymbol R}_3)}{\partial Z^2},
	}_{\rm vertex 3} 
	\right\}^{T}\;\;,
\end{eqnarray}
and 
\begin{eqnarray}
&&	{Y}_i = \nonumber\\
&&	 \left\{
\underbrace{y({\boldsymbol R}_1), 
	\frac{\partial y({\boldsymbol R}_1)}{\partial \xi}, 
	\frac{\partial y({\boldsymbol R}_1)}{\partial \eta},
	\frac{\partial^2 y({\boldsymbol R}_1)}{\partial \xi^2},  
	\frac{\partial^2 y({\boldsymbol R}_1)}{\partial \xi\partial \eta},
	\frac{\partial^2 y({\boldsymbol R}_1)}{\partial \eta^2},
}_{\rm vertex 1} \right. \nonumber \\
&&	\underbrace{y({\boldsymbol R}_2), 
	\frac{\partial y({\boldsymbol R}_2)}{\partial \xi}, 
	\frac{\partial y({\boldsymbol R}_2)}{\partial \eta},
	\frac{\partial^2 y({\boldsymbol R}_2)}{\partial \xi^2},  
	\frac{\partial^2 y({\boldsymbol R}_2)}{\partial \xi\partial \eta},
	\frac{\partial^2 y({\boldsymbol R}_2)}{\partial \eta^2},
}_{\rm vertex 2} \nonumber \\
&&
\left.
\underbrace{y({\boldsymbol R}_3), 
	\frac{\partial y({\boldsymbol R}_3)}{\partial \xi}, 
	\frac{\partial y({\boldsymbol R}_3)}{\partial \eta},
	\frac{\partial^2 y({\boldsymbol R}_3)}{\partial \xi^2},  
	\frac{\partial^2 y({\boldsymbol R}_3)}{\partial \xi\partial \eta},
	\frac{\partial^2 y({\boldsymbol R}_3)}{\partial \eta^2},
}_{\rm  vertex 3} 
\right\}^{T}\;\;.
\end{eqnarray}
the transformation between $Y_i$ and $\hat{Y}_i$ or between $\Lambda_i$ and $\hat{\Lambda}_i$  is obtained as follows,
\begin{eqnarray}
    Y_i=\bar{\bar T}\hat{Y}_i \;\;, \;\;
    \Lambda_i=\hat{\Lambda}_i \bar{\bar T} ^{-1}\;\;, 
\end{eqnarray}
where 
\begin{eqnarray}
	\bar{\bar T} = 
	\begin{bmatrix}
		1 & & & & & \\
		  & J_{11} & J_{12} & & & \\
		  & J_{21} & J_{22} & & & \\
		  &   &   & J_{11}^2 & 2J_{11}J_{12} & J_{12}^2 \\
		  &   &   & J_{11}J_{21} & J_{12}J_{21}+J_{11}J_{22} & J_{12}J_{22} \\
		  &   &   & J_{21}^2 & 2J_{12}J_{22} & J_{22}^2 \\
	\end{bmatrix}\;\;, \\
 J_{11}=\frac{\partial R}{\partial\xi}\;\;,\;\;
 J_{12}=\frac{\partial Z}{\partial\xi}\;\;,\;\;
 J_{21}=\frac{\partial R}{\partial\eta}\;\;,\;\;
 J_{22}=\frac{\partial Z}{\partial\eta}\;\;.
\end{eqnarray}

For the three-dimensional variable $Y(R,\phi,Z)$, the triangular C1 element is used in the $R,Z$ directions while cubic spline basis functions are adopted in the $\phi$ direction, written as follows,
\begin{eqnarray}
	Y(R,\phi,Z) &=& \sum_{i=1}^{6N_{\rm vert}}\sum_{j=1}^{N_\phi} \hat\Lambda_i(\xi,\eta) N_j(\phi) \hat{Y}_{i,j}  \;\;,
\end{eqnarray}
where the cubic spline basis function is defined as follows,
\begin{equation}
  N_{\rm{cubic}}(x) =
    \begin{cases}
       4/3+2x+x^2+x^3/6 \;\;, & \text{if $x\in[-2,-1)$}\\
       2/3-x^2-x^3/2 \;\;,    & \text{if $x\in[-1,0) $}\\
       2/3-x^2+x^3/2 \;\;,    & \text{if $x\in[0,1) $} \\
       4/3-2x+x^2-x^3/6 \;\;. & \text{if $x\in[1,2) $}
    \end{cases}       
\end{equation}
$N_j(x)$ and $N_{\rm cubic}(\phi)$ are related according to $N_{\rm cubic}(\phi)=N_j(x=(\phi-\phi_i)/\Delta\phi))$, where $\phi_i$ is the toroidal location of $i$th grid point, $\Delta\phi$ is the uniform grid size in the toroidal direction. 
In the following sections, $\Lambda_i$ and $Y_{i,j}$ will not be involved but we only use $\hat\Lambda_i$ and $\hat Y_{i,j}$. For the sake of simplicity,  we will omit the `$\hat{\;}$' in $\hat{\Lambda}_i$ and $\hat{Y}_{ij}$.  
 
\subsection{Weak form of field equations with charge/current deposition (projection)}
All field equations in our model can be written in a general form as follows,
\begin{eqnarray}
  L(R,\phi,Z)Y(R,\phi,Z)=b(R,\phi,Z) \;\;,
\end{eqnarray}
where $L$ is a linear differential operator. The weak form can be written as
\begin{eqnarray}
    \int {\rm d}R\, {\rm d}Z\, {\rm d}\phi\,\, S(R,\phi,Z) \Lambda_{i}(R,Z)N_{j}(\phi) 
    L(R,\phi,Z)Y(R,\phi,Z)
    \nonumber \\
     =\int {\rm d} R\, {\rm d}Z\, {\rm d}\phi\, S(R,\phi,Z) \Lambda_{i}(R,Z)N_{j}(\phi)
    b(R,\phi,Z) \;\;,
\end{eqnarray}
where $S(R,\phi,Z)$ is a weight function chosen to be the Jacobian $S=J=\{\nabla R\times\nabla\phi\times\nabla Z\}^{-1}=R$ in this work. 
The weak form of the quasi-neutrality equation, Amp\`ere's law, the iterative equation, and Ohm's law are
\begin{eqnarray}
\label{eq:mat_poisson}
    {\rm (\ref{eq:poisson_normalized})} \Rightarrow
    \bar{\bar{M}}_{\mathrm{P},L,ii',jj'} \cdot\delta\Phi_{i'j'}
    &&= C_{\rm P} \int {\rm d}R{\rm d}Z{\rm d}\phi\, J\delta N^{i,j}\tilde{N}_{ij} \;\;,  \\
\label{eq:mat_ampere0}
    {\rm (\ref{eq:ampere_normalize0})} \Rightarrow 
    \bar{\bar{M}}_{\mathrm{A},L,ii',jj'} \cdot\delta A^{\rm{h}}_{i'j',I=0} 
    &&= \bar{\bar{M}}_{\mathrm{A},R,ii',jj'} \cdot\delta A^{\rm{s}}_{i'j'}+ C_{\mathrm A} \delta J^{i,j}_\| \;\;,  \\
\label{eq:mat_ampere1}
    {\rm (\ref{eq:ampere_normalize1})}\Rightarrow
    \bar{\bar{M}}_{\mathrm{it},L,ii',jj'} \cdot\delta A^{h}_{i'j',I+1} 
    &&= \bar{\bar{M}}_{\mathrm{it},R,ii',jj'} \cdot\delta A^{h}_{i'j',I} 
    +\sum_s \frac{1}{d_{s}^2}\overline{\langle\delta A^{h}_{i,j,I} \rangle} \;, \\
\label{eq:mat_ohm}
    {\rm (\ref{eq:ohm_law0})} \Rightarrow
    \bar{\bar{M}}_{\mathrm{Ohm},L,ii',jj'} \cdot\partial_t \delta A^{\rm{s}}_{i'j'} 
    &&= 
    \bar{\bar{M}}_{\mathrm{Ohm},R,ii',jj'} \cdot \delta\Phi_{i'j'} \;\;.
\end{eqnarray}
{Equations (\ref{eq:mat_poisson})--(\ref{eq:mat_ohm}) are solved numerically which provides the perturbed fields for solving the gyro center's equations of motion. The Runge-Kutta fourth-order integration is adopted. 
The matrices and the terms on the right-hand side are as follows,
\begin{eqnarray}
    \bar{\bar{M}}_{\mathrm{P},L,ii',jj'}
    &&=-\sum n_{0s} \bar{m}_s \frac{B^2_{\rm{ref}}}{B^2}\int {\rm d} R\, {\rm d}Z\, {\rm d}\phi\, J  \nabla_\perp\Tilde{N}_{ij} 
    \cdot\nabla_\perp \Tilde{N}_{i'j'}\;\;,
    \nonumber \\ 
    \delta N^{i,j}
    &&=
    -\sum \Bar{q}_s\int {\rm d} R\, {\rm d} Z\, {\rm d}\phi\, J \delta\bar{N}_s(R,\phi,Z) \Tilde{N}_{ij} \;\;,\nonumber
\end{eqnarray}
\begin{eqnarray}
    \bar{\bar{M}}_{\mathrm{A},L,ii',jj'}
    &&=\int {\rm d} R\, {\rm d}Z\, {\rm d}\phi\, J
    \left[
    -\nabla_\perp\Tilde{N}_{ij} 
    \cdot\nabla_\perp \Tilde{N}_{i'j'}
    -\sum_s\frac{\bar{q}_s^2}{\bar m_s}C_{\mathrm A} 
    \Tilde{N}_{ij}\Tilde{N}_{i'j'} \right] \nonumber\;\;,
\end{eqnarray}
\begin{eqnarray}
    \bar{\bar{M}}_{\mathrm{A},R,ii',jj'}
    &&=\int {\rm d} R\, {\rm d}Z\, {\rm d}\phi\, J
    \nabla_\perp\Tilde{N}_{ij} 
    \cdot\nabla_\perp \Tilde{N}_{i'j'}\;\;,
    \nonumber\\
    \delta J^{i,j}_\|
    &&= 
    - \int {\rm d} R\, {\rm d} Z\, {\rm d}\phi\, J \delta\bar J_\|(R,\phi,Z) \Tilde{N}_{ij} 
 \;\;,\nonumber\\
    \bar{\bar{M}}_{\mathrm{it},L,ii',jj'}
    &&=
    \bar{\bar{M}}_{\mathrm{A},L,ii',jj'} \;\;, \nonumber \\
    \overline{\langle\delta A^{h}_{i,j,I} \rangle}  
    &&=
    \int {\rm d}R\,{\rm d}Z\,{\rm d}\phi\, J\overline{\langle\delta A^{h}_{I} \rangle} \tilde{N}_{ij}
    \;\;,\\
    \bar{\bar{M}}_{\mathrm{it},R,ii',jj'}
    &&=
    -C_{\mathrm A} \sum_s\frac{\bar{q}_s^2}{\bar m_s}\int {\rm d} R\, {\rm d} Z\, {\rm d}\phi\, J 
    \Tilde{N}_{ij}\Tilde{N}_{i'j'} \;\;,\nonumber \\
     \bar{\bar{M}}_{{\rm Ohm},L,ii',jj'} 
     &&=
    \int {\rm d} R\, {\rm d} Z\, {\rm d}\phi\, J 
    \Tilde{N}_{ij}\Tilde{N}_{i'j'}\;\;,
    \nonumber \\
     \bar{\bar{M}}_{{\rm Ohm},R,ii',jj'} 
     &&=
    -\int {\rm d} R\, {\rm d} Z\, {\rm d}\phi\, J 
    \Tilde{N}_{ij} \partial_\|\Tilde{N}_{i'j'} \;\;,
    \nonumber 
\end{eqnarray}
where $\Tilde{N}_{ij}=\Lambda_i(R,Z)N_j(\phi)$, $\overline{\langle\delta A^{h}_{ij,I} \rangle}$ is derived from $\delta A^{h}_{ij,I}$ by first interpolating the values of $\delta A^{h}_{ij,I}$ on markers (with or without gyro-average depending on the choice of the model), and then calculating the projection to the finite element basis. }

\subsection[Strong]{Gyro centers' equations of motion with the strong form in $\partial_t\delta{A}_\|^{\rm s}+\partial_\|\delta\Phi$}

\label{subsec:strong_AP}
Theoretically, since the ideal Ohm's law Eq.~(\ref{eq:ohm_law0}) is adopted, the term in the gyro centers' equations of motion can be simplified by making use of the identity $\partial_\| \delta\Phi + \partial_t\delta A_\|^{\rm s}=0$. However, numerically, the coefficients of the basis functions are solved for $\delta\Phi$ in Eq.~(\ref{eq:mat_poisson}) and $\delta{A}^{\rm s}$ in Eq.~(\ref{eq:mat_ohm}). Namely, the weak form of $\partial_\| \delta\Phi + \partial_t\delta A_\|^{\rm s}=0$ is satisfied but at an arbitrary location, $\partial_\| \delta\Phi + \partial_t\delta A_\|^{\rm s}$ is not exactly zero but finite. As a result, when calculating the gyro centers' equations of motion, both $\partial_\| \delta\Phi$ and $\partial_t\delta A_\|^{\rm s}$ are kept although these two terms almost cancel with each other numerically. This treatment is consistent since the coefficient of $\delta A_\|^{\rm s}$ in Eq.~(\ref{eq:mat_ampere0}) is the part we extract from $\delta A_\|$ numerically in the implementation. As a result, we use the numerical representation of $\partial_\|\delta\Phi$ and $\delta A^{\rm s}_\|$ in the gyro centers' equations of motion with the following form without explicit simplification $\partial_\| \delta\Phi + \partial_t\delta A_\|^{\rm s}=0$,
\begin{eqnarray}
	\delta \Phi(R,\phi,Z) &=& \sum_{i=1}^{6N_{\rm vert}}\sum_{j=1}^{N_\phi} \delta\Phi_{i,j}\Lambda_i(\xi,\eta) N_j(\phi)   \;\;, \\
	\delta A^{\rm s}_\|(R,\phi,Z) &=& \sum_{i=1}^{6N_{\rm vert}}\sum_{j=1}^{N_\phi}  \delta A^{\rm s}_{\|,i,j} \Lambda_i(\xi,\eta) N_j(\phi)  \;\;.
\end{eqnarray}

\section{Simulation setup and  results}
\label{sec:results}
Two typical cases are chosen to test TRIMEG-C1 in numerical studies of electromagnetic problems, including the Toroidal Alfv\'en Eigenmode (TAE) and the electromagnetic Ion Temperature Gradient (ITG) mode. The unstructured meshes are generated for the ITPA-TAE case \cite{konies2018benchmark} and the AUG case \cite{lauber2018strongly,lu2019development}, as shown in Fig. \ref{fig:meshes}.
For the ITPA-TAE case, the magnetic equilibrium is featured by the circular concentric magnetic surfaces. For the AUG case, the magnetic surfaces are elongated and up-down asymmetric. 
While sparser grids are shown in Fig. \ref{fig:meshes}, denser grids are adopted in the following simulations. Other parameters will be introduced in the following sections. Note that in this work, we focus on the formulation and the first set of numerical results highlighting the basic features of this scheme using the $\delta f$ method. The simulations are limited to the linear physics since the nonlinear simulations are more costly, which requires more computational resources and the optimization of the code in, e.g., GPU offloading and noise mitigation for the nonlinear simulations in the future. 

\begin{figure}[htbp]\centering
    \centering
    \includegraphics[width=0.4\textwidth]{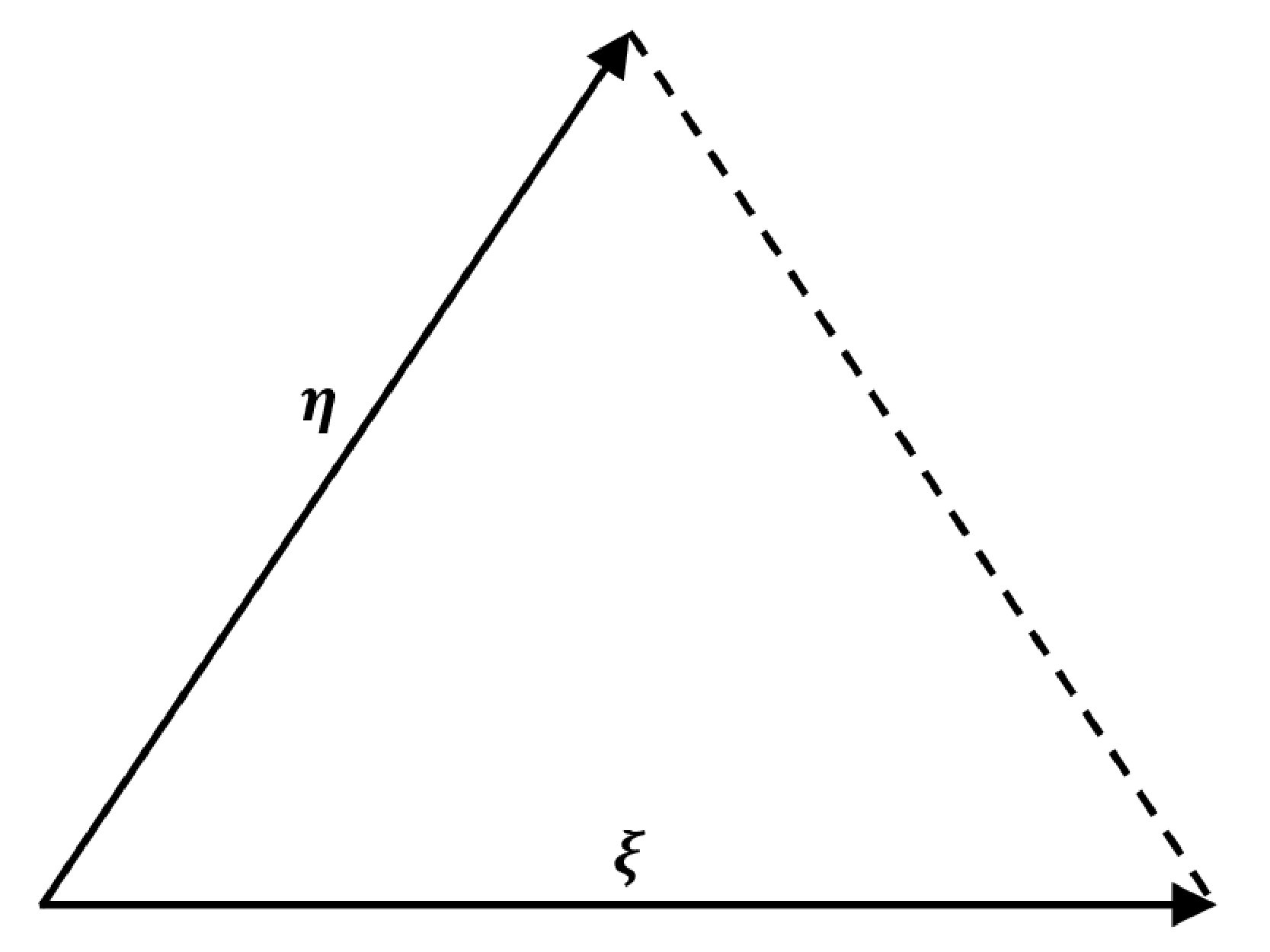}
    \caption{Local coordinates $(\xi,\eta)$ in a triangle.}
    \label{fig:triangle}
\end{figure}

\begin{figure}[htbp]\centering
    \centering
    \includegraphics[width=0.45\textwidth]{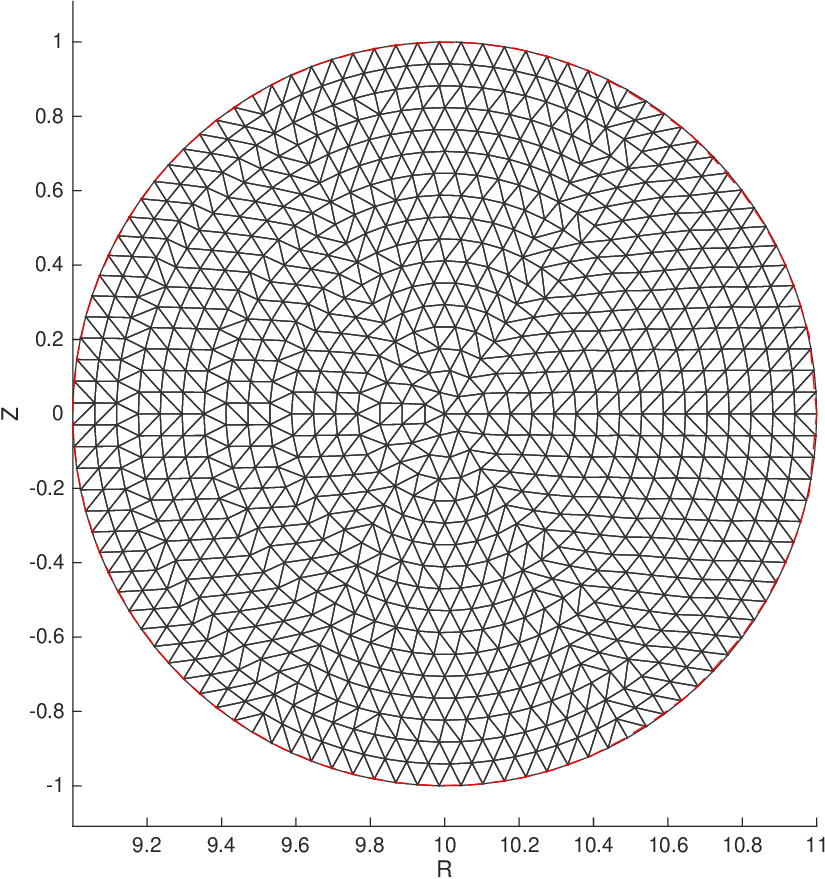}
    \includegraphics[width=0.45\textwidth]{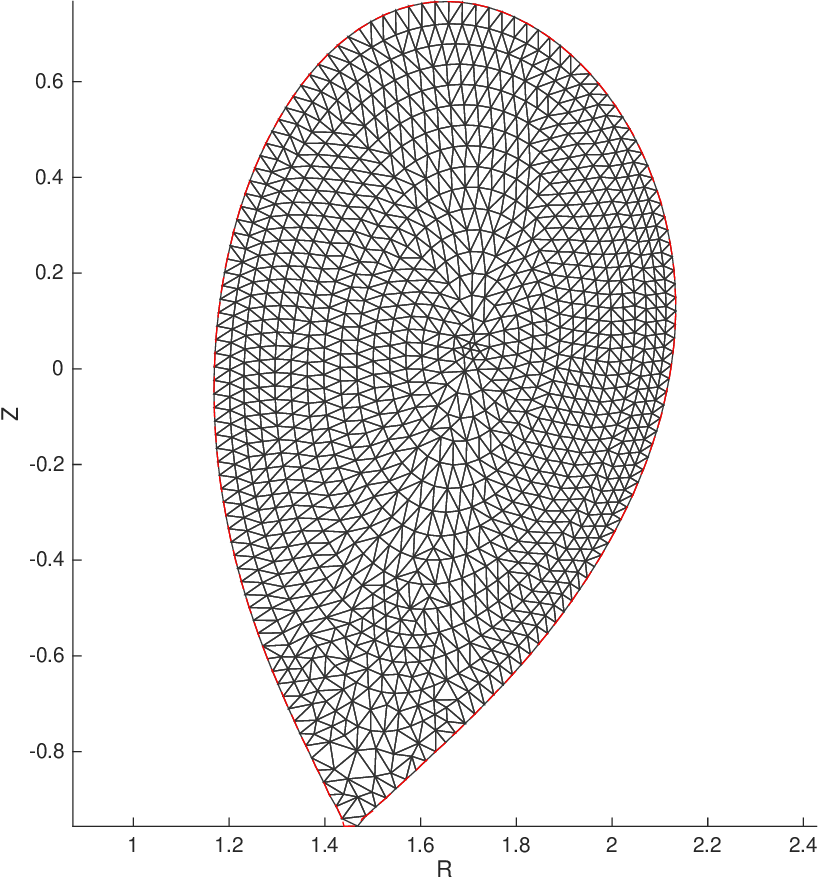}
    \caption{The unstructured meshes used for the ITPA-TAE case (left) and the AUG case (right).}
    \label{fig:meshes}
\end{figure}

\subsection{Numerical properties}
The vertices of the triangles are initialized from the axis to the edge, as shown in the left frame of Fig. \ref{fig:vertex_matrix} where the radial grid number is $N_r=6$ corresponding to 98 vertices (including 32 boundary vertices). The matrix nonzero elements of the quasi-neutrality equation are  shown in the right frame of Fig. \ref{fig:vertex_matrix}, where the grid number in the toroidal direction is $N_\phi=8$. The matrix structure is different than those using the structured meshes \cite{stier2024verification}. We use the sparse matrix to store the nonzero elements and the KSP iterative method in PETSc to solve the linear system \cite{balay2019petsc}. 

\begin{figure*}[htbp]\centering
    \centering
    \includegraphics[width=0.48\textwidth]{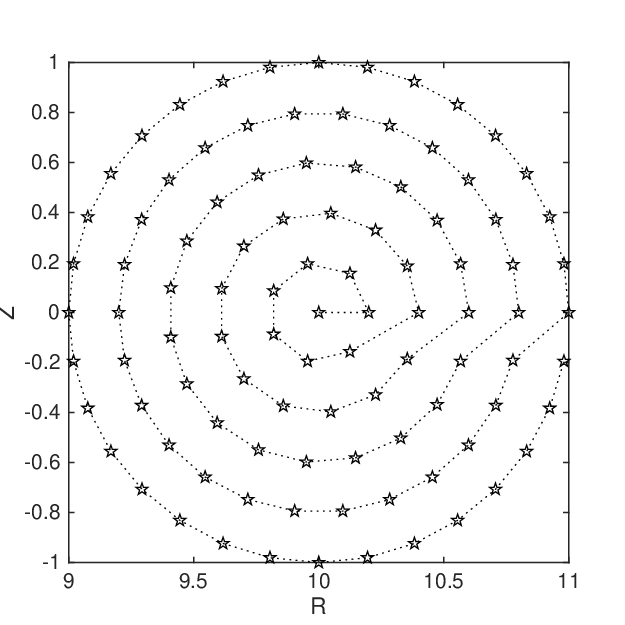}
    \includegraphics[width=0.48\textwidth]{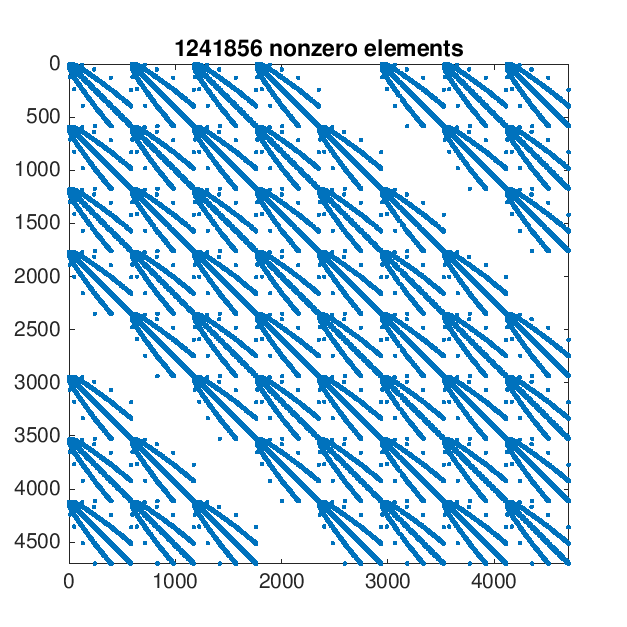}
    \caption{The vertices of the triangles (left) and the nonzero elements of the matrix of the quasi-neutrality equation (right).}
    \label{fig:vertex_matrix}
\end{figure*}

The accuracy of the C1 finite element scheme is tested using a two-dimensional solver for the elliptical equation with the analytical solution given as follows
\begin{eqnarray}
\label{eq:ms_eq}
    \nabla^2 f(R,Z)&=& S(R,Z) \;\;, \\
\label{eq:ms_rhs}
    S_{\rm ana}(R,Z)&=& 2\bar{Z}_0\bar{Z}_1\sin(k_R \bar{R}_0) \sin(k_Z \bar{Z}_0) \nonumber\\
                       &+&2 (\bar{R}_0+\bar{R}_1) \bar{Z}_0\bar{Z}_1k_R \cos(k_R \bar{R}_0) \sin(k_Z \bar{Z}_0) 
                       \nonumber\\
                       &-&\bar{R}_0 \bar{R}_1 \bar{Z}_0\bar{Z}_1k_R^2 \sin(k_R \bar{R}_0) \sin(k_Z \bar{Z}_0) \nonumber\\
                       &+&2 \bar{R}_0 \bar{R}_1 \sin(k_R \bar{R}_0) \sin(k_Z \bar{Z}_0) \nonumber\\
                       &+&2 (\bar{Z}_0+\bar{Z}_1) \bar{R}_0 \bar{R}_1 k_Z \sin(k_R \bar{R}_0) \cos(k_Z \bar{Z}_0) \nonumber\\
                       &-&\bar{R}_0 \bar{R}_1 \bar{Z}_0\bar{Z}_1k_Z^2 \sin(k_R \bar{R}_0) \sin(k_Z\bar{Z}_0);\;\;,\\
\label{eq:ms_sol}
    f_{\rm ana}(R,Z)&=&\bar{R}_0\bar{R}_1\bar{Z}_0\bar{Z}_1 \nonumber\\
    &\times&\sin(k_R\bar{R}_0)\sin(k_Z\bar{Z}_0)\;\;,
\end{eqnarray}
where the simulation domain is rectangular in $(R,Z)$ space $R_0\le R\le R_1$, $Z_0\le Z\le Z_1$, $\bar{R}_0=R-R_0$, $\bar{R}_1=R-R_1$, $\bar{Z}_0=Z-Z_0$, $\bar{Z}_1=Z-Z_1$. 
The C0 and the C1 finite element schemes are implemented in Matlab to solve Eq.~\ref{eq:ms_eq} with $S_{\rm ana}$ given in Eq.~\ref{eq:ms_rhs}. The numerical solution is $f_{\rm nu}$. The relative error is calculated as 
\begin{eqnarray}
    \epsilon_{\rm nu}=\sqrt{\frac{\sum_{i,j} (f_{{\rm ana},i,j}-f_{{\rm nu},i,j})^2}{\sum_{i,j} f_{{\rm ana},i,j}^2}} \;\;,
\end{eqnarray}
where $i,j$ are the indices in $R$, $Z$ directions, respectively. The relative errors are shown in Fig. \ref{fig:c0c1compare}. The error using C1 method is significantly lower than the C0 method. In addition, as the grid number increases in both directions, the scaling of the accuracy of the C1 scheme is significantly better than that of the C0 scheme. This result motivates us to implement the C1 scheme in TRIMEG-C1 in Fortran. The C0 scheme is not implemented in Fortran for the one-to-one comparison in the self-consistent electromagnetic particle simulations. 

\begin{figure}[htbp]\centering
    \centering
    \includegraphics[width=0.85\textwidth]{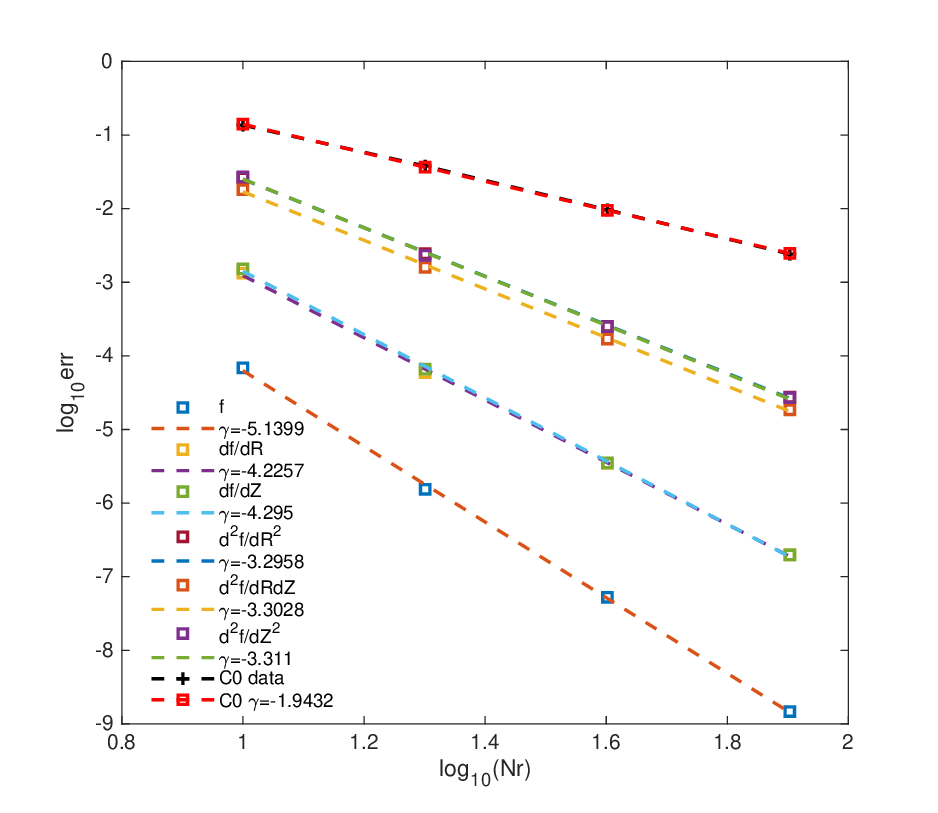}
    \caption{The convergence of the C0 scheme and the C1 scheme. The red squares give the convergence order of the C0 scheme $\gamma=-1.9432$. The blue squares give the convergence order in the zeroth order derivative of the solution of the C1 scheme $\gamma=-5.1399$. The yellow and green squares give the convergence order in the first order derivative of the solution of the C1 scheme $\gamma=-4.2257$ for $\partial f/\partial R$ and $\gamma=-4.2950$ for $\partial f/\partial Z$. The other three types of squares give  the convergence order in the second order derivative of the solution of the C1 scheme $\gamma=-3.2958$ for $\partial^2 f/\partial R^2$, $\gamma=-3.3028$ for $\partial^2 f/\partial R\partial Z$ and $\gamma=-3.3110$ for $\partial^2 f/\partial Z^2$.}
    \label{fig:c0c1compare}
\end{figure}

The iterative Amp\`ere solver in Eq.~(\ref{eq:ampere_iterative}) is tested for evaluating
the accuracy of $\delta A_\|^{\rm h}$. We show the convergence of the iterative Amp\`ere solver of a typical run in Fig. \ref{fig:ampere_iterative}. Good convergence is observed for the base case ($T_{\rm EP} = 400$ keV), with the radial grid number $N_r=64$. The correction in $\delta A_\|^{\rm h}$ is smaller as the number of iterations increases. In the initial state, the convergence is
better than at later times since the marker distribution deviates away from the Maxwell distribution due to the finite orbit width effect and mirror force, which leads to a larger discrepancy of the $\delta A_{\|,0}^{\rm h}$ from the rigorous solution  $\sum_{I=0,1,2,\ldots}\delta A_{\|,I}^{\rm h}$.
Nevertheless, the convergence is good, and the correction to  $\delta A_\|^{\rm h}$ is suppressed to be lower than $0.5\%$ in only 3 iterations.

\begin{figure}[htbp]\centering
    \centering
    \includegraphics[width=0.485\textwidth]{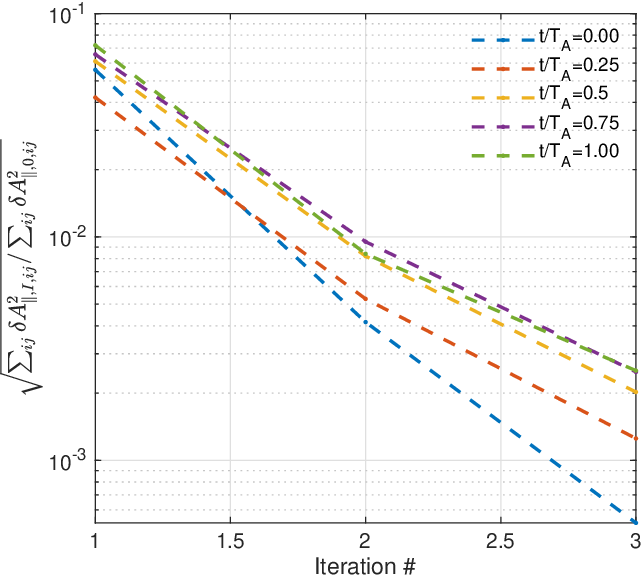}
    \caption{The correction in $\delta A_\|^{\rm h}$ versus the iteration number, which indicates the convergence of the iterative Amp\`ere solver. In the first iteration, the correction in $\delta A_\|^{\rm h}$ is $3\%\sim8\%$. In the third iteration, the correction is lower than $0.5\%$. For the Alfv\'enic modes, we choose $T_A$ as the time unit defined as $T_A=v_A(r=0)/[2q(r=r_{\rm c})R_0]$ which is close to the period of the Toroidal Alfv\'en eigenmode. }
    \label{fig:ampere_iterative}
\end{figure}

\subsection{ITPA-TAE case}
\label{subsec:itpatae}
The toroidicity induced Alfv\'en eigenmode driven by energetic particles is simulated using the parameters defined by the ITPA group \cite{konies2018benchmark}. 
The major radius $R_0=10 \; {\rm m}$, minor radius $a=1$ m, on-axis magnetic field $B_0=3$ T, and the safety factor profile $q(r)=1.71+0.16r^2$. The electron density and temperature are constant with $n_{{\rm e}0}=2.0\times10^{19}\;\rm{m}^{-3}$, $T_{\rm{e}}=1$~keV. The ratio of the electron pressure to the magnetic pressure is $\beta_{\rm e}\approx 9\times 10^{-4}$. The Larmor radius of the thermal {ions} is $\rho_{\rm{ti}}=m_{\rm{i}}v_{\rm{ti}}/(eB_{\rm{axis}})=1.52\times10^{-3}$ m. If the nominal mass ratio $m_{\rm i}/m_{\rm e}=1836$ is used, the ratio between the adiabatic part ($\delta A_\|^{\rm h}/d_{\rm e}^2$) and the non-adiabatic part ($\nabla_\perp^2\delta A^{\rm{h}}$) in the left-hand side of Amp\`ere's equation is $1/(d_{\rm e}^2k_\perp^2)\approx \beta_{\rm e}/(k_\perp\rho_{\rm{ti}})^2(m_{\rm{i}}T_{\rm i}/m_{\rm{e}}T_{\rm{e}})\approx1.622\times10^3$, where $k_\perp\approx nq/r=6\times1.75/0.5=21$.  This ITPA-TAE case is featured with a small electron skin depth ($d_{\rm e}\approx1.182\times10^{-3}$ m) and suffers from the ``cancellation problem'' if the pullback scheme is not adopted. While the nominal mass ratio has been adopted in our previous work \cite{lu2023full}, in the following, we use $m_{\rm i}/m_{\rm e}=100$ to reduce the computational cost. Note that the poloidal Fourier filter has not been used in our work to study edge physics in the future.  Consequently, the time step size ($\Delta t$) is smaller (the maximum $\Delta t\propto k_\|v_A$, where $k_\|$ is the parallel wave vector in the system, $v_A$ is the Alfv\'en velocity) than that with structured meshes and the Fourier filter \cite{lu2023full} by a factor of $\sim1/160$. With $m_{\rm i}/m_{\rm e}=100$, the time step size can be increased by $\sim 4$ times. Correspondingly, $1/(d_{\rm e}^2k_\perp^2)\approx88.34$, which is still an appropriate case with moderate to small electron skin depth for testing the cancellation problem.  
 
 The EP density profile is given by 
\begin{eqnarray}
\label{eq:nEP1d}
	n_{\rm{EP}}(r)&=&n_{\rm{EP},0}c_3 \exp\left[ -\frac{c_2}{c_1} \tanh\left(\frac{{r}-c_0}{c_2}\right)\right]\;\;, \\
	\frac{{\rm d}\ln n_{\rm{EP}}}{{\rm d} r} &=&\cosh^{-2}\left(\frac{ r-c_0}{c_2}\right)\;\;,
\end{eqnarray}
where the normalized radial-like coordinate $r=\sqrt{(\psi-\psi_{\rm axis})/(\psi_{\rm edge}-\psi_{\rm axis})}$, $n_{\rm{EP},0} =1.44131\times10^{17}\; \rm{m}^{-3}$, the subscript `$\rm{EP}$' indicates EPs (energetic particles), $c_0 = 0.491 23$, $c_1 =0.298 228$, $c_2 =0.198 739$, $c_3 =0.521 298$. The EP temperature is $400\; {\rm keV}$ for the base case. The $n=6$ mode is simulated by applying a toroidal Fourier filter. The initial perturbation in the marker weight is applied with two poloidal harmonics with $m=10, 11$. $16\times10^6$ electron markers, $4\times10^6$ ion markers and $4\times10^6$ energetic particle markers are simulated. The radial grid number is $N_r=32$ and 3 iterations are taken in the iterative Amp\`ere solver leading to $<2.5\%$ correction in $\delta A_\|^{\rm h}$ in the last iteration. 
The Finite Larmor Radius effect is omitted by eliminating the gyro-average operation.  The simulation is run on 64 nodes (Intel Xeon IceLake-SP 8360Y) of the supper computer Raven, with 72 CPU cores on each node with the Processor Base Frequency $2.4$ GHz and the Max Turbo Frequency $3.5$ GHz. It takes $\sim5$ hours to simulate one TAE period ($1 ~T_A$). 

The time evolution of the total field energy is shown in Fig. \ref{fig:energy1d_itpa_400kev}. The total field energy is defined as 
\begin{eqnarray}
    E_\Phi=-C_{\rm P}\int {\rm d}V \frac{1}{G_{\rm P}}\delta\bar\Phi\delta \bar N \;\;, \;\;
    E_{\rm A}=-C_{\rm A}\int {\rm d}V \delta \bar A_\|\delta \bar J_\| \;\;,
\end{eqnarray}
where $G_{\rm P}$ is defined in Eq.~\ref{eq:poisson_normalized} and $E_{\rm \Phi}$ is an approximate value of the field energy in $\delta\Phi$ in the limit $|k_\|/k_\perp|\ll 1$  and  $|\nabla_\perp\ln G_{\rm P}|/|\nabla\ln\delta\bar\Phi|\ll1$.   
The ratio $E_{\rm A}/E_\Phi\rightarrow1$ as the exponential growth stage appears, which is consistent with the feature of the Toroidal Alfv\'en mode. Since the initial perturbation is close to the physics solution dominated by the $m=10, 11$ harmonics, the physics solution appears quickly in only 2 TAE periods. 
The two-dimensional structures of $\delta \Phi$ and $\delta A_\|$ are shown in Fig. \ref{fig:itpa_mode2dAP_Tf400}. The ballooning structure of $\delta \Phi$ and the anti-ballooning structure of $\delta A_\|$ are consistent with the previous results using the structured meshes \cite{lu2023full,konies2018benchmark}. More detailed studies, including extensive parameter scans, will require more computationally expensive runs and further optimization of computational efficiency in the future.

\begin{figure}[htbp]\centering
    \centering
    \includegraphics[width=0.425\textwidth]{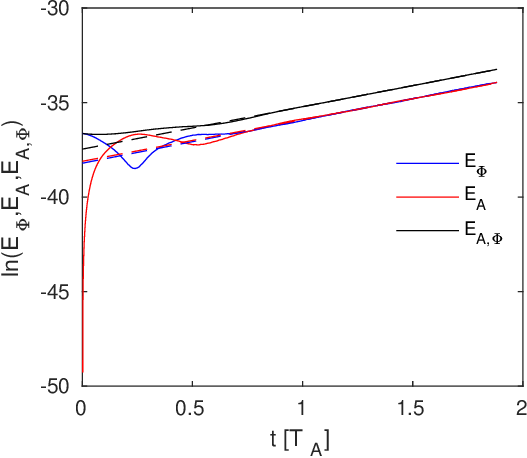}
    \caption{Time evolution of the total field energy for $T_{\rm EP}=400$ keV. $E_{\rm \Phi}$ and $E_{\rm A}$ are the total energy in $\delta\Phi$ and $\delta A_\|$ respectively. $E_{\rm A,\Phi}=E_{\rm A}+E_{\rm\Phi}$. }
    \label{fig:energy1d_itpa_400kev}
\end{figure}

\begin{figure}[htbp]\centering
    \centering
    \includegraphics[width=0.85\textwidth]{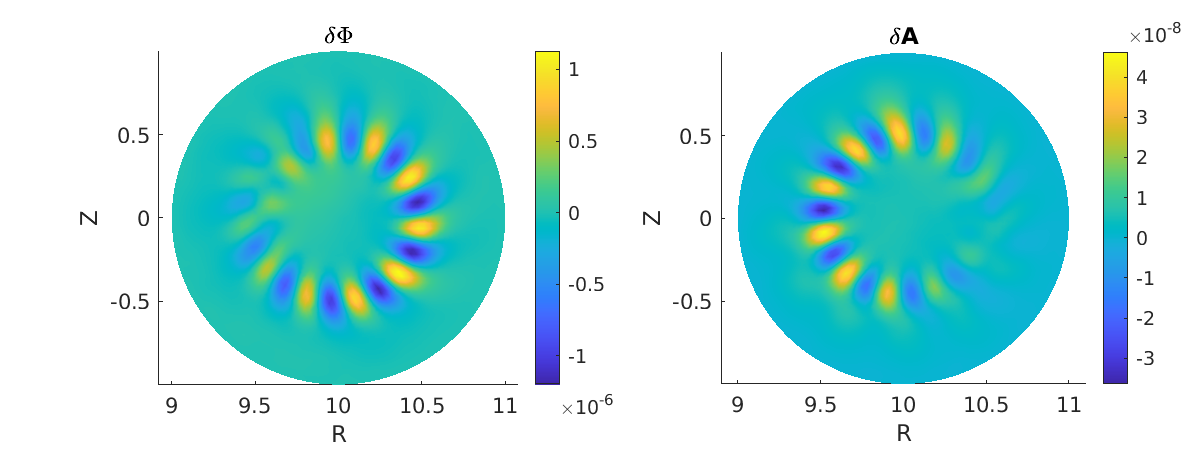}
    \caption{2D mode structures of $\delta\Phi$ and $\delta{A}_\|$ for $T_{\rm EP}=400$ keV in the end of the simulation. }
    \label{fig:itpa_mode2dAP_Tf400}
\end{figure}


\subsection{AUG case for the ion temperature gradient mode studies}
In this section, we use the realistic geometry of the
ASDEX Upgrade (AUG) case with shot number 34924 at 3.600 s. This is a typical discharge for the studies of energetic particle physics in different codes and models \cite{lauber2018strongly} and ion temperature gradient (ITG) mode in the TRIMEG code \cite{lu2019development}. In the simulations, we use the experimental equilibrium but analytical density and temperature proﬁles. The parameters are chosen for the studies of the ITG mode. 
The temperature and density profiles indicated by $A(r)$ and the corresponding normalized logarithmic gradients indicated by $L_{\rm ref}/L_A$ are given by
\begin{eqnarray}
\label{eq:nT}
\frac{A(r)}{A(r_0)}&=&\exp\left[-\kappa_A W_A\frac{a}{L_{\rm ref} }\tanh\left( \frac{r-r_{\rm c}}{W_A a} \right) \right]\;\;,\\
\label{eq:dlnnTdr}
\frac{L_{\rm ref}}{L_A}&=&-L_{\rm ref} \frac{{\rm d}\ln A}{{\rm d}r}=\kappa_A\cosh^{-2} \left( \frac{r-r_{\rm c}}{W_A a} \right) \;\;,
\end{eqnarray}
where the subscript `c' denotes the center of the gradient profile and  $r_{\rm c}=0.5$, $W_A=0.3$
with the normalized radial coordinate defined in Section \ref{subsec:itpatae}. 
This study aims to test the capability of treating realistic geometry with minimum technical complexity. The fully self-consistent treatment of the density/temperature proﬁle and the equilibrium will be addressed in another work.

The nominal values of the two basic parameters using the on-axis density and temperature are $\beta_{\rm N}=3.1544\%$ corresponding to $\beta_{\rm axis}=0.525\%$ since $B_{\rm axis}=2.4506$ T, $\rho_{\rm N}=5.5240\times10^{-3}$ m. In the following tests, $\rho_{\rm N}$ is increased to make the simulation less costly. From our simulations using $\rho_{\rm N}=0.04, 0.02, 0.01$, the computational cost to avoid the numerical crash depends on several physics parameters as follows,
\begin{eqnarray}
    C_{\rm comp}\propto 
    C_{\Delta t}C_{\rm marker} C_{m_e} C_\beta
    \approx \rho_{\rm N}^{-1} \rho_{\rm N}^{-2} m_e^{-1/2} \beta^{-1/2}\;\;,    
\end{eqnarray}
where $C_{\Delta t}$ is due to the reduction of the time step size $\Delta t$ as $\rho_{\rm N}$ decreases, $C_{\rm marker}$ is due to the increment of the marker number as $\rho_{\rm N}$ decreases, $C_{m_e}$ and $C_{\beta}$ are due to the reduction of the time step size $\Delta t$ as $m_e/\beta$ decreases. 
The $n=4$ mode is simulated using $4\times10^6$ ions and $10^6$ electrons for $\rho_{\rm N}=0.02$ m. 2 iterations are taken in the iterative Amp\`ere solver leading to $<0.1\%$ correction in $\delta A_\|^{\rm h}$ in the last iteration. The simulation is run on 12 nodes (AMD Genoa EPYC 9354) of the TOK cluster, with 32 CPU cores on each node,  with the Processor Base Frequency $3.25$ GHz and the Max Turbo Frequency $3.8$ GHz. It takes 0.32 hours to simulate one normalized time unit $t_{\rm N}=R_{\rm N}/v_{\rm N}$. 

The time step size is $\Delta t=0.016\, t_{\rm N}$. The white noise is initialized in the marker weights according to the function $w_p=R_{\rm noi}A_{\rm noi}S_{\rm noi}$ where $S_{\rm noi}=\exp[-(r_p-r_{\rm c,noi})^2/r_{\rm w,noi}^2]$, $A_{\rm noi}=10^{-5}$, $r_{\rm c,noi}=0.55$, $r_{\rm w,noi}=0.2$ and $R_{\rm noi}\in(0,1)$ is a random number from the uniform distribution. The ITG mode is destabilized. The time evolution of the total field energy is shown in Fig. \ref{fig:energy1d_aug_core}.
The two-dimensional structures of $\delta \Phi$ and $\delta A_\|$ are shown in Fig. \ref{fig:aug_mode2dAP_core}. The ballooning structure of $\delta \Phi$ is observed and is quantitatively consistent with the previous result using ORB5 \cite{lanti2020orb5} or the electrostatic version of TRIMEG-C0 \cite{lu2019development}.
The tilting of the 2D mode structure shows the symmetry-breaking properties of the gyrokinetic solution due to the symmetry-breaking mechanisms such as the up-down asymmetry of the equilibrium, the profile shearing, and the global effect, and a more detailed analysis will be performed by closer comparisons with other codes and analyses \cite{hornsby2018global,lu2017symmetry}. 
In the future, we will perform more comparisons with other gyrokinetic simulations and studies with open field lines for the whole plasma studies. 

\begin{figure}[htbp]\centering
    \centering
    \includegraphics[width=0.425\textwidth]{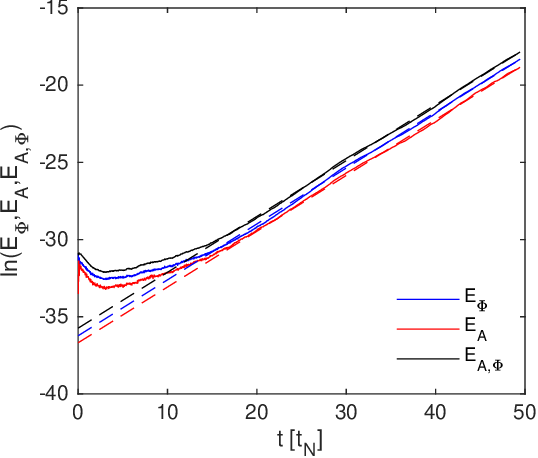}
    \caption{Time evolution of the total field energy for the AUG case with $\beta_{\rm N}=3.1544\%$, $\rho_{\rm N}=0.02$ m. The definition of $E_{\rm A}$, $E_{\rm \Phi}$ and $E_{\rm A,\Phi}$ is the same as those in Fig.~\ref{fig:energy1d_itpa_400kev}. }
    \label{fig:energy1d_aug_core}
\end{figure}

\begin{figure}[htbp]\centering
    \centering
    \includegraphics[width=0.85\textwidth]{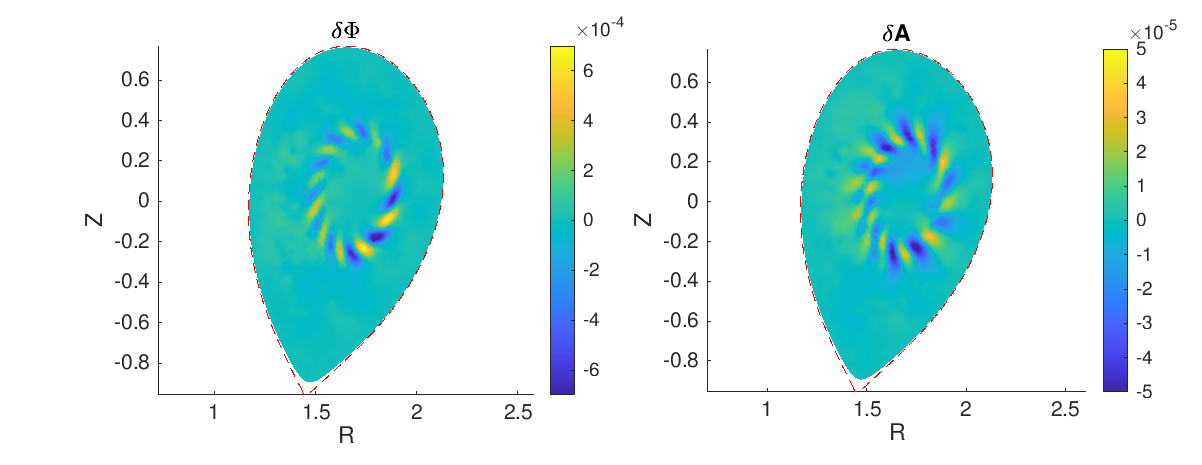}
    \caption{2D mode structures of $\delta\Phi$ and $\delta{A}_\|$ in the end of the simulation for the AUG case with $\beta_{\rm N}=3.1544\%$, $\rho_{\rm N}=0.02$ m.}
    \label{fig:aug_mode2dAP_core}
\end{figure}

\section{Conclusions and outlook}
In this work, we have developed the scheme for the gyrokinetic electromagnetic particle simulations in triangular meshes with C1 finite elements in the TRIMEG-C1 code. The mixed variable/pullback scheme has been implemented for gyrokinetic electromagnetic particle simulations. The filter-free treatment in the poloidal cross-section with triangular meshes is adopted, which allows future application in the open field line region. The correction of the finite $\delta E_\|^{\rm s}\equiv -\partial_t\delta A^{\rm s}_\|-\partial_\|\delta\Phi$ due to the finite element representation in the gyro centers' equation of motion (in Section \ref{subsec:strong_AP}), namely, the numerical residual of the nonzeros $\delta E_\|^{\rm s}$,  is considered and is important in improving the simulation quality, especially when the poloidal filter is not applied. 

The numerical results using the ITPA-TAE parameters validate our implementation, showing its capability in moderate to small electron skin depth regimes. Simulations using experimental parameters confirm its applicability in
realistic plasma geometry. The high-order C1 finite element scheme for particle simulations developed in this work provides high accuracy for physics studies, opens up the possibilities of nonlinear gyrokinetic simulations in the TRIMEG code including the whole spectrum (multiple toroidal harmonics), and the inclusion of an open field line regime as well as the transport studies in constants of motion space with multi-species including energetic particles \cite{meng2024energetic}. Further improvement in the computational efficiency using more advanced noise reduction schemes and GPU accelaration are needed in the future for nonlinear simulations using experimental parameters.

\begin{acknowledgments}
Z.X. Lu appreciates the fruitful discussions with Ralf Kleiber, Alberto Bottino, Weixing Wang and Laurent Villard. The simulations in this work were run on TOK cluster, MPCDF and CINECA (MARCONI) supercomputers. 
The Eurofusion projects TSVV-8, ACH/MPG, TSVV-10 and ATEP are acknowledged. 
This work has been carried out within the framework of the EUROfusion Consortium, funded
by the European Union via the Euratom Research and Training Programme (Grant Agreement No
101052200—EUROfusion). Views and opinions expressed are however those of the author(s)
only and do not necessarily reflect those of the European Union or the European Commission.
Neither the European Union nor the European Commission can be held responsible for them.
\end{acknowledgments}

\appendix

\section{The complete form of the gyro centers' equations of motion}
\label{subsec:gcmotion_RZ_complete}
In this section, all variables such as $m_s$ are normalized. In $(R,\phi,Z)$, the unperturbed variable ${\boldsymbol b}^*_0$ is as follows,
\begin{eqnarray}
    b_{0R}^* &=& b_R-\frac{B_{\rm ref}}{B_\|^*}\rho_{\rm N}v_\|\partial_Zb_\phi \;\;,\\
    b_{0Z}^*&=&b_Z+\frac{B_{\rm ref}}{B_\|^*}\rho_{\rm N}v_\|\left(
    \partial_Rb_\phi+\frac{b_\phi}{R} \right) \;\;,\\
    b_{0\phi}^*&=&b_\phi+\frac{B_{\rm ref}}{B_\|^*}\rho_{\rm N}v_\|\left(
    \partial_Zb_R-\partial_Rb_Z \right) \;\;,
\end{eqnarray}
and the complete form is
\begin{eqnarray}    
    b_R^* &=& b_{0R}^*
    +\frac{B_{\rm ref}}{B_\|^*}\rho_{\rm N}\left[\frac{b_Z}{R}\partial_\phi-b_\phi\partial_Z\right]\langle\delta A_\|^{\rm s}\rangle\;\;, \\
    b_Z^* &=& b_{0Z}^*
    +\frac{B_{\rm ref}}{B_\|^*}\rho_{\rm N}
    \left[b_\phi\partial_R-\frac{b_R}{R}\partial_\phi\right]\langle\delta A_\|^{\rm s}\rangle\;\;, \\
    b_\phi^* &=& b_{0\phi}^*
    +\frac{B_{\rm ref}}{B_\|^*}\rho_{\rm N}
    \left[
    b_R\partial_Z-b_Z\partial_R\right]\langle\delta A_\|^{\rm s}\rangle\;\;.
\end{eqnarray}
Based on Eqs.~(\ref{eq:dR0dt_normalized})--(\ref{eq:du0dt_normalized}), the equilibrium part of the gyro center motion is as follows,
\begin{eqnarray}
    \dot R_0 &=& b^*_{0R} u_\|
        +C_{\rm d} B_\phi\partial_ZB\;\;, \\
    \dot Z_0 &=& b_{0Z}^* u_\|
        -C_{\rm d}B_\phi\partial_R B\;\;, \\
    \dot\phi_0 &=& \frac{b_{0\phi}^*}{R} u_\|
        +\frac{C_{\rm d}}{R}(B_Z\partial_RB-B_R\partial_ZB)  \\
    \dot u_{\|,0} &=& 
    -\mu (b_{0R}^*\partial_RB+b_{0Z}^*\partial_ZB) \;\;.
\end{eqnarray}
where $C_{\rm d}=(m_s/q_s)\rho_{\rm N}\mu B B_{\rm ref}/(B^2B^*_\|)$, $\delta G=\delta\Phi-u_\|\delta A_\|$.
Based on Eqs.~(\ref{eq:d1Rdt_normalized})--(\ref{eq:du1dt_normalized}), the perturbed part of the equations of motion is
\begin{eqnarray}
    \delta\dot{R} &=& C_{\rm E}\left(
        b_\phi\partial_Z\delta G-\frac{1}{R}b_Z\partial_\phi\delta G\right)
        -\frac{\bar q_s}{\bar m_s}\langle\delta\bar A_\|^{\rm h}\rangle {b}^*_R \;\;, \\
    \delta\dot{Z} &=& C_{\rm E} \left(
        -b_\phi\partial_R\delta G+\frac{1}{R}b_R\partial_\phi\delta G\right)
        -\frac{\bar q_s}{\bar m_s}\langle\delta\bar A_\|^{\rm h}\rangle {b}^*_Z \;\;, \\
    \delta\dot{\phi} &=& \frac{C_{\rm E}}{R}\left(
        b_Z\partial\delta_R G-b_R\partial_Z\delta G\right) 
        -\frac{1}{R}\frac{\bar q_s}{\bar m_s}\langle\delta\bar A_\|^{\rm h}\rangle {b}^*_\phi\;\;, \\
    \delta\dot{u}_\| &=& -\frac{q_s}{m_s}\left[ 
          b_R^*(\partial_R\delta\Phi-u_\|\partial_R\delta A^{\rm h}_\|) 
        + b_Z^*(\partial_Z\delta\Phi-u_\|\partial_Z\delta A^{\rm h}_\|) \right. \nonumber\\
        &&+
        \frac{b_\phi^*}{R}(\partial_\phi\delta\Phi-u_\|\partial_\phi\delta A^{\rm h}) 
        +\partial_t\delta A_\|^{\rm s} \nonumber\\
       && + \left.
       \frac{m_s}{q_s}(\dot{R}_{\mu}\partial_R+\dot{Z}_{\mu}\partial_Z+\dot{\phi}_{\mu}\partial_\phi)\langle\delta A_\|^{\rm s}\rangle 
    \right] \;\;,
\end{eqnarray}
where $C_{\rm E}=\rho_{\rm N}B_{\rm ref}/B^*_\|$, $C_\mu=(m_s/q_s)\rho_{\rm N}\mu BB_{\rm ref}/(B^2B^*_\|)$, $\dot{R}_{\mu}=C_\mu B_\phi\partial_ZB$, $\dot{Z}_{\mu} =-C_\mu B_\phi\partial_RB$, $\dot{\phi}_{\mu} =\frac{C_\mu}{R}( B_Z\partial_RB-B_R\partial_ZB)$. 
If ideal Ohm's law for $\delta A^{\rm s}_\|$ in Eq.~(\ref{eq:ohm_law0}) is used, we obtain
\begin{eqnarray}
      \dot u_{\|,1} &=& -\frac{q_s}{m_s} \left[ 
        - b_Ru_\|\partial_R\delta A^{\rm h}_\| 
        - b_Zu_\|\partial_Z\delta A^{\rm h}_\|
        - ({b_\phi}/{R}) u_\|\partial_\phi\delta A^{\rm h}_\|
        \right. \nonumber\\
        &&+\left.\Delta b_R^*\partial_R\delta\Phi
          +\Delta b_Z^*\partial_Z\delta\Phi
          +\Delta b_\phi^*\partial_\phi\delta\Phi \right] \nonumber\\
        &&-(\dot{R}_{\mu}\partial_R+\dot{Z}_{\mu}\partial_Z+\dot{\phi}_{\mu}\partial_\phi)\langle\delta A_\|^{\rm s}\rangle\;\;,
\end{eqnarray}
where $\Delta{\boldsymbol b}^*={\boldsymbol b}^*-{\boldsymbol b}$. 

\section{Dominant terms in the gyro centers' equations of motion}
\label{subsec:gcmotion_RZ_simplified}
The dominant terms in the gyro centers' equations of motion are obtained by replacing ${\boldsymbol b}^*$ with ${\boldsymbol b}$ but still keeping the magnetic drift due to the magnetic curvature. The equilibrium part of the GC motion in $(R,\phi,Z)$ is as follows,
\begin{eqnarray}
    \dot R_0 &=& b_R u_\|
        +C_{\rm d} B_\phi\partial_ZB\;\;, \\
    \dot Z_0 &=& b_Z u_\|
        -C_{\rm d}B_\phi\partial_R B\;\;, \\
    \dot\phi_0 &=& \frac{b_\phi}{R} u_\|
        +\frac{C_{\rm d}}{R}(B_Z\partial_RB-B_R\partial_ZB)\;\;,  \\
    \dot u_{\|,0} &=& 
    -\mu (b_{0R}^*\partial_RB+b_{0Z}^*\partial_ZB) \;\;.
\end{eqnarray}
where $C_{\rm d}=(m_s/q_s)\rho_{\rm N}(v_\|^2+\mu B)B_{\rm ref}/B^3$, $\delta G=\delta\Phi-u_\|\delta A_\|$.
The perturbed part of the equations of motion is
\begin{eqnarray}
    \delta \dot R &=& C_{\rm E}\left(
        b_\phi\partial_Z\delta G-\frac{1}{R}b_Z\partial_\phi\delta G\right)
        -\frac{\bar q_s}{\bar m_s}\langle\delta\bar A_\|^{\rm h}\rangle {b}_R \;\;, \\
    \delta \dot Z &=& C_{\rm E} \left(
        -b_\phi\partial_R\delta G+\frac{1}{R}b_R\partial_\phi\delta G\right)
        -\frac{\bar q_s}{\bar m_s}\langle\delta\bar A_\|^{\rm h}\rangle {b}_Z \;\;, \\
    \delta \dot\phi &=& \frac{C_{\rm E}}{R}\left(
        b_Z\partial\delta_R G-b_R\partial_Z\delta G\right)
        -\frac{1}{R}\frac{\bar q_s}{\bar m_s}\langle\delta\bar A_\|^{\rm h}\rangle {b}_\phi\;\;, \\
    \delta \dot u_{\|} &=& -\frac{q_s}{m_s}\left[ 
          b_R(\partial_R\delta\Phi-v_\|\partial_R\delta A^{\rm h}_\|) 
        + b_Z(\partial_Z\delta\Phi-v_\|\partial_Z\delta A^{\rm h}_\|) \right. \nonumber\\
        &&+
        \frac{b_\phi}{R}(\partial_\phi\delta\Phi-v_\|\partial_\phi\delta A^{\rm h}_\|) 
        +\partial_t\delta A_\|^{\rm s} \nonumber\\
       && + \left.
       \frac{m_s}{q_s}(\dot{R}_{\mu}\partial_R+\dot{Z}_{\mu}\partial_Z+\dot{\phi}_{\mu}\partial_\phi)\langle\delta A_\|^{\rm s}\rangle 
    \right] \;\;,   
\end{eqnarray}
where $C_{\rm E}=\rho_{\rm N}B_{\rm ref}/B$, $C_\mu=(m_s/q_s)\rho_{\rm N}\mu BB_{\rm ref}/B^3$, $\dot{R}_{\mu}=C_\mu B_\phi\partial_ZB$, $\dot{Z}_{\mu} =-C_\mu B_\phi\partial_RB$, $\dot{\phi}_{\mu} =\frac{C_\mu}{R}( B_Z\partial_RB-B_R\partial_ZB)$. If ideal Ohm's law is used, 
\begin{eqnarray}
      \delta \dot v_{\|} &=& +\frac{q_s}{m_s} \left( 
          b_Rv_\|\partial_R\delta A^{\rm h}_\| 
        + b_Zv_\|\partial_Z\delta A^{\rm h}_\|
        + \frac{b_\phi}{R} v_\|\partial_\phi\delta A^{\rm h}_\|
        \right) \nonumber\\
        &&-(\dot{R}_{\mu}\partial_R+\dot{Z}_{\mu}\partial_Z+\dot{\phi}_{\mu}\partial_\phi)\langle\delta A_\|^{\rm s}\rangle\;\;.
\end{eqnarray}

\newpage
\providecommand{\newblock}{}


\begin{thebibliography}{10}
	\expandafter\ifx\csname url\endcsname\relax
	\def\url#1{{\tt #1}}\fi
	\expandafter\ifx\csname urlprefix\endcsname\relax\def\urlprefix{URL }\fi
	\providecommand{\eprint}[2][]{\url{#2}}
	
	\bibitem{lee1983gyrokinetic}
	Lee W 1983 {\em Phys. Fluids\/} {\bf 26} 556
	
	\bibitem{chen2007electromagnetic}
	Chen Y and Parker S~E 2007 {\em Journal of Computational Physics\/} {\bf 220}
	839--855
	
	\bibitem{hatzky2019reduction}
	Hatzky R, Kleiber R, K{\"o}nies A, Mishchenko A, Borchardt M, Bottino A and
	Sonnendr{\"u}cker E 2019 {\em Journal of Plasma Physics\/} {\bf 85} 905850112
	
	\bibitem{mishchenko2005gyrokinetic}
	Mishchenko A, K{\"o}nies A and Hatzky R 2005 Gyrokinetic simulations with a
	particle discretization of the field equations {\em Joint Varenna-Lausanne
		International Workshop on Theory of Fusion Plasmas\/} (Societa Italiana di
	Fisica) pp 315--322
	
	\bibitem{bao2018conservative}
	Bao J, Lin Z and Lu Z 2018 {\em Physics of Plasmas\/} {\bf 25}
	
	\bibitem{mishchenko2019pullback}
	Mishchenko A, Bottino A, Biancalani A, Hatzky R, Hayward-Schneider T, Ohana N,
	Lanti E, Brunner S, Villard L, Borchardt M {\em et~al.\/} 2019 {\em Computer
		Physics Communications\/} {\bf 238} 194--202
	
	\bibitem{kleiber2024euterpe}
	Kleiber R, Borchardt M, Hatzky R, K{\"o}nies A, Leyh H, Mishchenko A, Riemann
	J, Slaby C, Garc{\'\i}a-Rega{\~n}a J, Sanchez E {\em et~al.\/} 2024 {\em
		Computer Physics Communications\/} {\bf 295} 109013
	
	\bibitem{lu2021development}
	Lu Z, Meng G, Hoelzl M and Lauber P 2021 {\em Journal of Computational
		Physics\/} {\bf 440} 110384
	
	\bibitem{sturdevant2021verification}
	Sturdevant B~J, Ku S, Chac{\'o}n L, Chen Y, Hatch D, Cole M, Sharma A, Adams M,
	Chang C, Parker S {\em et~al.\/} 2021 {\em Physics of Plasmas\/} {\bf 28}
	072505
	
	\bibitem{lu2023full}
	Lu Z, Meng G, Hatzky R, Hoelzl M and Lauber P 2023 {\em Plasma Physics and
		Controlled Fusion\/} {\bf 65} 034004
	
	\bibitem{wang2015distinct}
	Wang W, Ethier S, Ren Y, Kaye S, Chen J, Startsev E and Lu Z 2015 {\em Nuclear
		Fusion\/} {\bf 55} 122001
	
	\bibitem{lin1998turbulent}
	Lin Z, Hahm T~S, Lee W, Tang W~M and White R~B 1998 {\em Science\/} {\bf 281}
	1835--1837
	
	\bibitem{chang2017fast}
	Chang C, Ku S, Tynan G, Hager R, Churchill R, Cziegler I, Greenwald M, Hubbard
	A and Hughes J 2017 {\em Phys. Rev. Lett.\/} {\bf 118} 175001
	
	\bibitem{lu2019development}
	Lu Z, Lauber P, Hayward-Schneider T, Bottino A and Hoelzl M 2019 {\em Phys.
		Plasmas\/} {\bf 26} 122503
	
	\bibitem{holzl2024non}
	H{\"o}lzl M, Huijsmans G, Artola F, Nardon E, Becoulet M, Schwarz N, Cathey A,
	Pamela S, Aleynikova K, Antlitz F {\em et~al.\/} 2024 {\em Nuclear Fusion\/}
	
	\bibitem{jardin2004triangular}
	Jardin S~C 2004 {\em Journal of Computational Physics\/} {\bf 200} 133--152
	
	\bibitem{jo2022development}
	Jo G, Kwon J~M, Seo J and Yoon E 2022 {\em Computer Physics Communications\/}
	{\bf 273} 108265
	
	\bibitem{lanti2020orb5}
	Lanti E, Ohana N, Tronko N, Hayward-Schneider T, Bottino A, McMillan B,
	Mishchenko A, Scheinberg A, Biancalani A, Angelino P {\em et~al.\/} 2020 {\em
		Computer Physics Communications\/} {\bf 251}
	
	\bibitem{rekhviashvili2023gyrokinetic}
	Rekhviashvili L, Lu Z, Hoelzl M, Bergmann A and Lauber P 2023 {\em Physics of
		Plasmas\/} {\bf 30}
	
	\bibitem{mishchenko2023global}
	Mishchenko A, Borchardt M, Hatzky R, Kleiber R, K{\"o}nies A, N{\"u}hrenberg C,
	Xanthopoulos P, Roberg-Clark G and Plunk G~G 2023 {\em Journal of Plasma
		Physics\/} {\bf 89} 955890304
	
	\bibitem{wang2015identification}
	Wang W, Ethier S, Ren Y, Kaye S, Chen J, Startsev E, Lu Z and Li Z 2015 {\em
		Physics of Plasmas\/} {\bf 22}
	
	\bibitem{lu2015intrinsic}
	Lu Z, Wang W, Diamond P, Tynan G, Ethier S, Gao C and Rice J 2015 {\em Physics
		of Plasmas\/} {\bf 22}
	
	\bibitem{lu2015effects}
	Lu Z, Wang W, Diamond P, Tynan G, Ethier S, Chen J, Gao C and Rice J 2015 {\em
		Nuclear Fusion\/} {\bf 55} 093012
	
	\bibitem{strang1971analysis}
	Strang G and Fix G~J 1971 {\em An analysis of the finite element method\/}
	
	\bibitem{konies2018benchmark}
	K{\"o}nies A, Briguglio S, Gorelenkov N, Feh{\'e}r T, Isaev M, Lauber P,
	Mishchenko A, Spong D, Todo Y, Cooper W {\em et~al.\/} 2018 {\em Nucl.
		Fusion\/} {\bf 58} 126027
	
	\bibitem{lauber2018strongly}
	Lauber P, Geiger B, Papp G, Guimarais L, Poloskei P~Z, Igochine V, Maraschek M,
	Pokol G, Hayward-Schneider T, Lu Z {\em et~al.\/} 2018 {\em proceedings of
		the 27th IAEA Fusion energy\/}
	
	\bibitem{stier2024verification}
	Stier A, Bottino A, Boesl M, Pinto M~C, Hayward-Schneider T, Coster D, Bergmann
	A, Murugappan M, Brunner S, Villard L {\em et~al.\/} 2024 {\em Computer
		Physics Communications\/} {\bf 299} 109155
	
	\bibitem{balay2019petsc}
	Balay S, Abhyankar S, Adams M, Brown J, Brune P, Buschelman K, Dalcin L, Dener
	A, Eijkhout V, Gropp W {\em et~al.\/} 2019 {\em PETSc users manual\/}
	
	\bibitem{hornsby2018global}
	Hornsby W, Angioni C, Lu Z, Fable E, Erofeev I, McDermott R, Medvedeva A,
	Lebschy A, Peeters A, Team A~U {\em et~al.\/} 2018 {\em Nuclear Fusion\/}
	{\bf 58} 056008
	
	\bibitem{lu2017symmetry}
	Lu Z, Fable E, Hornsby W, Angioni C, Bottino A, Lauber P and Zonca F 2017 {\em
		Physics of Plasmas\/} {\bf 24}
	
	\bibitem{meng2024energetic}
	Meng G, Lauber P, Lu Z, Bergmann A and Schneider M 2024 {\em Nuclear Fusion\/}
	{\bf 64} 096009
	
\end{thebibliography}
\end{document}